\documentclass[12pt]{article}                   
\usepackage[a4paper, margin=2.5cm]{geometry}    

\usepackage{setspace}   
\usepackage{parskip}    
\raggedright

\usepackage[none]{hyphenat}     
\usepackage{amsmath, amssymb}   
\usepackage{siunitx}            
\usepackage{graphicx}           
\usepackage{rotating}           
\usepackage{cite}               
\usepackage{authblk}            

\usepackage{hyperref}           
\hypersetup{
  colorlinks = true,
  linkcolor  = blue,
  filecolor  = blue,
  urlcolor   = blue,
  citecolor  = blue
}

\usepackage{fancyhdr}
\usepackage{lastpage}
\title{On the vortex transport and blade interactions in a reversible pump-turbine}

\author[1]{Chirag Trivedi}

\affil[1]{\small Waterpower Laboratory, Department of Energy and Process Engineering, NTNU — Norwegian University of Science and Technology, Alfred Getz' vei 4, 7491 Trondheim, Norway.}

\date{\small \today}

\begin{document}
\maketitle

\begin{abstract}
\singlespacing
\noindent Pumped storage type hydropower plants play an important role in mitigating real-time energy flexibility. Reversible pump-turbines undergo extreme operating conditions such as runaway and speed-no-load. Very limited studies are undertaken to understand the stochastic flow under these conditions in the reversible pump-turbine. The present study investigates the unsteady vortical flow, its transportation, and interaction with the blades at speed-no-load. Large eddy simulations are conducted in both turbine and pump modes. The computational domain contains 120 million nodes. Numerical results provided evidence of a large longitudinal vortex that develops on the high-pressure side of the blade, and transports into the blade passage and develops the unsteady ``string of swirls". The results also showed another ``string of swirls" in the draft tube, where flow in the center is reversible (pumping). The resulting flow instability is very high, and it has the potential to induce fatigue damage to the blades.
\end{abstract}

\section{Introduction}
\label{sec:into}
Hydropower is a valuable source of clean energy allowing real-time energy production. It provides essential flexibility and operates in tandem with solar and wind power generators \cite{Caralis2012}. Hydropower enables us to store energy up to the scale of several TWh depending on the reservoir capacity. The global energy production from hydropower is around 4 500 TWh annually by 2024. Pumped storage accounts for approximately 189 GW and aims to reach 1 500 GW by 2023. Reversible pump-turbine is an essential component of an energy storage type hydropower plant. The pump-turbine generates electricity in turbine mode and stores energy in pump mode operations depending on requirement. The pump-turbine plays a critical role in the energy system to mitigate flexibility. In pump mode, rotational direction is opposite and the water flows in reverse direction, i.e., from the lower reservoir to the upper reservoir. The pump-turbine is integrated with other valuable components, such as spiral casing to guide the water, stay vanes to distribute the water circumferentially, guide vanes to control the water flow through the turbine, runner to extract energy from the water, and draft turbine to recover the pressure energy. Modern pump-turbines have achieved an efficiency of up to 94\% in both turbine and pump mode operations at design load \cite{Stelzer1977}.

To provide essential ancillary services for real-time flexibility, the pump-turbines undergo extreme and highly damaging operating conditions, such as zero discharge, speed-no-load, turbine brake, pump brake, and runaway. Frequent operation at these conditions causes significant fatigue damage to the turbine components, specifically the rotating blades \cite{Zuo2016}. Stochastic turbulent flow in the turbine induces forced vortex, high amplitude vibrations, and the dynamic instability \cite{Zeng2016}. Turbine operation at these conditions is generally avoided. However, to mitigate real-time flexibility, operating quadrant change occurs from the generating to the pumping and vice versa; consequently, the pump-turbines undergo the damaging conditions.

The flow phenomenon in the blade channel can be categorized into three categories: (1) bulk flow detachment, (2) vortical and (3) stall. The first category corresponds to the bulk flow detachment in the channel. When the turbine is operated at stable load, around $\pm$ 20\% of the design load \cite{Olimstad2012a,Olimstad2012b,Magnoli2014}. The main flow is detached from the blade suction side due to the steep curvature of the blade; however, it remains streamlined throughout the channel \cite{Zobeiri2009}. A large intermittent vortex sets up seldom. This type of flow neither alters the blade loading nor induces dynamic instability, and broadly, the turbine operation remains stable. The second category corresponds to the local inception of vortical structures and remains detached from the nearby wall. This type of flow condition prevails in far off-design operations such as speed-no-load, runaway, turbine brake, pump brake \cite{Hasmatuchi2012}. The flow is permanently separated from the high- and low-pressure sides of the blade \cite{Wojcik2014}. This is because the flow leaves the guide vane at a low angle of attack, and it does not align with the blade inlet angle. The turbulent flow is, in fact, separated beyond the threshold point, where the possibility of rebounding on a no-slip wall is minimal. This results in different categories of vortex, such as horseshoe vortex, corner vortex, end (hub/shroud) wall vortex, and passage vortex \cite{Wu2001,Sharma1987,Schneider2014}. Furthermore, angular rotation (high tangential velocity) of the turbine runner provides momentum for the flow, causing the inception of local swirls. Yamamoto et al. \cite{Yamamoto2017} have studied the vortex characteristics in the blade channel through a novel visualization technique that provided optical access (including a borescope with a deflecting prism) to the runner blade channel.

The third category corresponds to the stall condition of the flow in the runner, referred to as s-curve instabilities and the rotating stall \cite{Binama2021,Xue2025,Xu2024,Wang2024,Shi2024,Liu2024}. The flow rate is small, and, on the other hand, the flow angle is very small, and the circumferential velocity is high. This results in an immediate separation of the flow at the leading edge, and continue along the blade channel. The turbine undergoes severe vibrations and other dynamic instabilities \cite{Wu2013,Dorfler2013}. During the stall condition, one or several blade and guide vane channels are partially blocked ($\pm 50\%$ of the nominal discharge), known as ``stalled cell", while the other channels comprise sound flow. The stalled cell(s) are rotating with a sub-synchronous frequency, generally 60 – 65\% of the runner rotation \cite{Cavazzini2016,Widmer2011}, and the stall propagates around 45 – 70\% of the runner rotation \cite{Xia2017}. The blade and guide vane channels have blockage of around ± 50\% of the nominal discharge during stall condition \cite{Pacot2016}. The majority of the instabilities have been studied through numerical simulations, and some experimentally. The numerical results, specifically pressure pulsations, are validated with experimental data. Botero et. al. \cite{Botero2014} and Hasmatuchi et al. \cite{Hasmatuchi2011} have investigated stall conditions experimentally. They used non-intrusive tufting techniques for the detection of rotating stall in a pump-turbine in vaneless space. Thin fluorescent wires (diameter = \SI{0.14}{mm}) are fixed at the mid-span of one guide vane channel towards the vaneless space. During the normal flow condition, the wires align themselves in the flow direction and do not disturb the flow. High-speed imaging (10,000 fps) was performed through a Plexiglas window located at the turbine diffuser.

Pumped storage type hydropower plants play an important role in mitigating real-time energy flexibility. The reversible pump-turbines undergo extreme operating conditions. The vortex induced vibrations cause significant fatigue damage to the turbine blades. Very limited or no studies are undertaken to understand the flow characteristics at these extreme conditions in the reversible pump-turbine. Furthermore, to the best of the author’s knowledge, no investigations are conducted at speed-no-load condition of a pump-turbine, neither turbine mode, nor pump mode. The flow field in the blade channel is significantly different from that of the turbine mode. The present study aims to investigate the unsteady vortical flow, its inceptions and transportation, and interaction with the rotating blades of the reversible pump-turbine at speed-no-load. The study encapsulates both modes of operation, i.e., Pump and turbine.

\section{Reversible pump-turbine}
\label{sec:rpt}
This study is conducted on a reversible pump-turbine, and the turbine is a reduced scale model (1:5.1) of a prototype. The reversible pump-turbine includes 6 blades, 28 guide vanes, 14 stay vanes, spiral casing and elbow type draft tube. \autoref{tab:turbine_parameters} shows the overall parameters of the turbine at the design load. The inlet and outlet diameters of the model runner are \SI{0.631}{m} and \SI{0.349}{m}, respectively.

\begin{table}[htb]
\caption{Overall design data of the investigated pump-turbine in generating mode}
\label{tab:turbine_parameters}
\vspace{1em}
\centering
    \begin{tabular}{l l}
        \hline
        \textbf{Parameter} & \textbf{Value} \\
        \hline
        Runner inlet diameter ($D_1$) & \SI{0.631}{\meter} \\
        Runner outlet diameter ($D_2$) & \SI{0.349}{\meter} \\
        Runner inlet height ($B_1$) & \SI{0.059}{\meter} \\
        Runner specific speed ($N_q$) & \SI{27.1}{\meter\tothe{3/4} \second\tothe{-1/2}} \\
        Angular speed ($n^{*}$) & \SI{10.8}{\per\second} \\
        Flow rate ($Q^{*}$) & \SI{0.275}{\cubic\meter\per\second} \\
        Head ($H^{*}$) & \SI{29.3}{\meter} \\
        Inlet blade angle ($\beta_1$) & \ang{12} \\
        Outlet blade angle ($\beta_2$) & \ang{12.8} \\
        Guide vane angle ($\alpha^{*}$) & \ang{10} \\
        Speed factor $(N_{ED})$ & 0.223 \\
        Discharge factor $(Q_{ED})$ & 0.133 \\
        \hline
    \end{tabular}
\end{table}

\begin{figure}[htbp]
    \centering
    \includegraphics[width=1\linewidth]{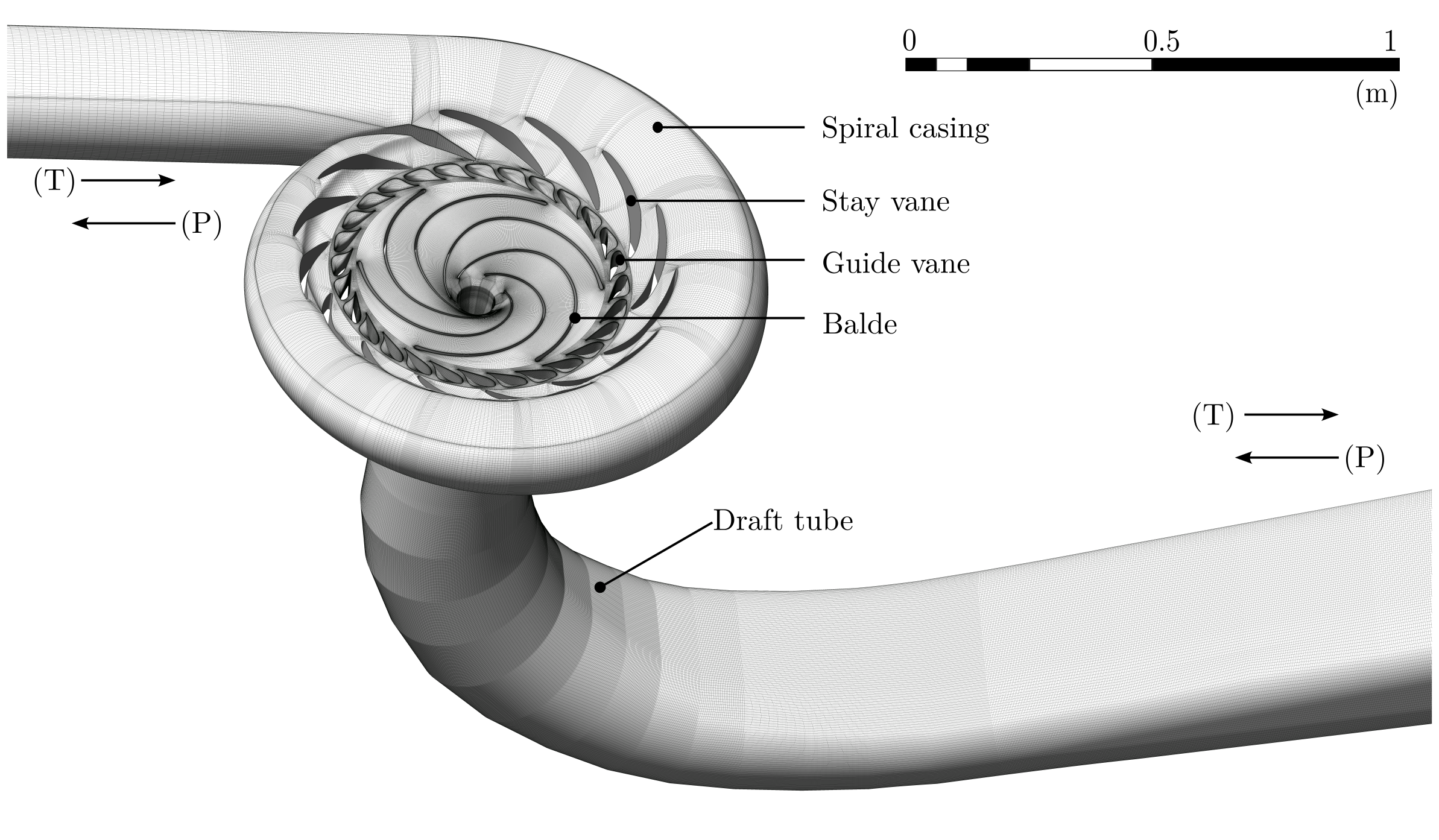}
    \caption{Computational model of the investigated pump-turbine. The flow directions (P) and (T) indicate the pump and turbine mode, respectively.}
    \label{fig:Figure1}
\end{figure}

The computational model of the turbine with all components was considered. \autoref{fig:Figure1} shows the computational model of the final mesh. Hexahedral mesh was created in all components. ANSYS\textsuperscript{\textregistered} ICEM CFD\textsuperscript{\texttrademark} was used for creating the hexahedral mesh in the spiral casing and draft tube domains. ANSYS\textsuperscript{\textregistered} TGrid\textsuperscript{\texttrademark} was used for creating the hexahedral mesh in the guide vane and the runner domains. The simulations were carried out using ANSYS\textsuperscript{\textregistered} CFX\textsuperscript{\texttrademark}. A high-performance computing cluster available at the Norwegian University of Science and Technology was used for simulations. Steady state simulations were conducted during the first phase. Reynolds-averaged Navier-Stokes (RANS) based shear stress transport (SST) turbulence model was used \cite{Menter2009}. The results of SST were used to carry out a mesh verification study. RANS modeling approaches do not provide essential information of the unsteady vortex due to averaging of the fluctuating component and minimum resolution of the turbulent length scales \cite{Hanjalic2005,Keck2008,Spalart2015}. Scale-adaptive simulation (SAS-SST) is a hybrid approach, taking the benefit of fine mesh and dynamically adjusting the resolved length scales, Artificial Forcing function, and adjusting the von Karman length scale. Detailed information about the mathematical model of SAS-SST is available in the literature \cite{Menter2006_SAS,Menter2010_SAS_AF}. SAS tends to be an explicit transfer of kinetic energy, where the flow instabilities are not at the level of LES requirement. SAS uses forcing terms in the momentum equation to convert into resolved kinetic energy from the modeled kinetic energy \cite{Menter2010_SAS_AF}.

\begin{equation}
Q_{\mathrm{SAS}} = 
\max \Bigg[
  \rho \, \zeta_2 \, S^{2} 
  \left( \frac{L}{L_{\mathrm{vk}}} \right)^{2}
  -
  C_{\mathrm{SAS}}
  \left( 
    \frac{2 \rho k}{\sigma_{\phi}}
  \right)
  \max \left(
    \frac{1}{k^{2}} 
    \frac{\partial k}{\partial x_j}
    \frac{\partial k}{\partial x_j},
    \,
    \frac{1}{\omega^{2}}
    \frac{\partial \omega}{\partial x_j}
    \frac{\partial \omega}{\partial x_j},
    \,
    0
  \right)
\Bigg]
\end{equation}

\noindent where $\zeta_2 = 1.47, \sigma_{\phi} = \frac{2}{3}, C_{\mathrm{SAS}} = 2.$ Another turbulence modeling approach is the Detached Eddy Simulation (DES) that provides hybrid solutions by combining the elements of RANS and LES formulations. RANS is applied inside attached and partially separated boundary layers, and the LES is applied at the largely separated regions \cite{Strelets2001}. This approach enables switching from SST-RANS model to the LES in regions where the turbulent length $(L_t)$ ) predicted by the RANS model is larger than the local mesh spacing. In this case, the length scale used in the computation of the dissipation rate in the equation for the turbulent kinetic energy is replaced by the local mesh spacing $(\Delta)$.

\begin{equation}
    \varepsilon 
= \beta^{*} k \omega
= \frac{k^{3/2}}{L_{t}}
\;\longrightarrow\;
k^{3/2}\,(E_{\mathrm{DES}}\,\Delta)
\qquad \text{for } C_{\mathrm{DES}} \Delta < L_{t}
\end{equation}

\begin{equation}
    \Delta = \max\!\left( \Delta_i \right),
\qquad
L_t = \frac{\sqrt{k}}{\beta^{*}\,\omega}.
\end{equation}

The final simulations were conducted using large eddy simulation (LES). Mesh in the runner domain refined significantly following the recommendations provided by Choi and Moin \cite{Choi2012}. LES enables a separation between large and small scales. The governing equations for LES are obtained by filtering the time-dependent Navier-Stokes equations in physical space. The filtering process effectively filters out the eddies whose scales are smaller than the filter width or mesh spacing used in the computations. A filtered variable is denoted in the following by an overbar and is defined by,

\begin{equation}
    \bar{\Phi}(x)
= \int \Phi(x') \, G(x; x') \, dx'.
\end{equation}

where, $G$ is the filter function that determines the scale of the resolved eddies. The unresolved part can be defined as,

\begin{equation}
    \Phi' = \Phi - \bar{\Phi}.
\end{equation}

For resolving all turbulent length scales, LES is a natural choice \cite{Menter2021}. This is highly expensive and time-consuming. The wall-adapted local eddy-viscosity (WALE) model provides balanced solutions for the LES. This is an algebraic model, and it produces almost no eddy-viscosity in wall bounded flow. This allows building of reliable laminar to turbulent transition layer. Furthermore, WALE is designed to return the correct wall-asymptotic $y^+$ variation of the sub-grid scale viscosity. It offers therefore the same advantages as the Dynamic Smagorinsky-Lilly model, and at the same time does not require an explicit (secondary) filtering. The wall-adapted local eddy-viscosity (WALE) model \cite{Nicoud1999} is formulated locally and uses the following equation to compute the eddy-viscosity.

\begin{equation}
    \mu_{sgs}
=
\rho \left( C_w \, \Delta \right)^{2}
\frac{
    \left( S_{ij}^{d} S_{ij}^{d} \right)^{3/2}
}{
    \left( \bar{S}_{ij} \bar{S}_{ij} \right)^{5/2}
    +
    \left( S_{ij}^{d} S_{ij}^{d} \right)^{5/4}
}.
\end{equation}

\begin{equation}
    S_{ij}^{d}
=
\bar{S}_{ik}\bar{S}_{kj}
+
\bar{\Omega}_{ik}\bar{\Omega}_{kj}
-
\frac{1}{3}\,\delta_{ij}
\left( 
    \bar{S}_{mn}\bar{S}_{mn}
    -
    \bar{\Omega}_{mn}\bar{\Omega}_{mn}
\right).
\end{equation}

\begin{equation}
    \bar{\Omega}_{ij}
=
\frac{1}{2}
\left(
\frac{\partial \bar{U}_i}{\partial x_j}
-
\frac{\partial \bar{U}_j}{\partial x_i}
\right).
\end{equation}

The General Grid Interface (GGI) between the rotating and stationary domains was modeled as a transient rotor-stator interface. This enables the flow variables to exchange information at every time-step and update the angular position of the rotating domain. This interface requires extra simulation time as compared to the frozen rotor interface type. The inlet and outlet boundary conditions were mass flow inlet and pressure outlet types for both pump and turbine mode operations. Available experimental data \cite{Kerlefsen2025} are used to prescribe the boundary condition. The inlet and outlet boundaries are swapped for pump mode operation. The inlet boundary corresponds to the inlet of the draft tube, and the outlet boundary corresponds to the outlet of the spiral casing. Rotational direction of the runner was also changed for pump mode operation.

The simulations were performed at speed-no-load condition (9\% guide vane opening) of the reversible pump-turbine. Investigations are conducted in both turbine and pump modes. \autoref{tab:mode_parameters} shows flow properties obtained through experiments at speed-no-load conditions. The turbine load of 9\% is the minimum sustainable load in turbine mode and pump mode. The design point for the turbine mode and pump mode corresponds to the 100\% load. The pressure$p_1$ and $p_2$ indicate the inlet and outlet static pressure in turbine and pump modes, respectively.

\begin{table}[hbt]
\caption{Flow properties at speed-no-load condition in turbine and pump modes operation}
\label{tab:mode_parameters}
\vspace{1em}
\centering
    \begin{tabular}{l l l l l l l}
        \hline
        Mode 
        & {$n$ (\si{\per\second})}
        & {$Q$ (\si{\cubic\meter\per\second})}
        & {$p_1$ (\si{\pascal})}
        & {$p_2$ (\si{\pascal})}
        & {$H$ (\si{\meter})}
        & {$T$ (\si{\newton\meter})} \\
        \hline
        Turbine (T) & 261 & 0.025 & 207560 & 80974  & 11.9 & 52  \\
        Pump (P)    & 410 & 0.024 & 94345  & 152410 & 11.9 & 158 \\
        \hline
    \end{tabular}
\end{table}

\section{Verification and validation}
\label{sec:vv}
Verification and validation of the computational domain were carried out before proceeding with the LES simulations. The verification was carried out using methods presented in the literature for hydraulic turbine \cite{Trivedi2019}. The grid convergence method (GCI) is used for the estimation and reporting of uncertainty due to discretization \cite{Roache1997}. This method is primarily based on the Richardson extrapolation method \cite{Richardson1911}; in addition, it takes advantage of the extrapolated mesh and the variable values. The verification and validation are conducted at the best efficiency point of the turbine. The global parameter, toque, was considered for numerical verification and validation. Torque value represents the global order of accuracy and the important performance parameter of the turbine. It indicates the performance of both stationary and rotating domains, spiral casing, guide vanes, runner and draft tube. The apparent order $(k)$ is 1.6, which is within the discretization scheme, high resolution scheme $(0 < \beta < 1)$, where $\beta = 0$ is the first order and $\beta = 1$ is the second order upwind scheme. The numerical uncertainty in the fine-grid (40.56 million nodes) solution for the torque value is 3.54\%. More information on the GCI analysis and the verification is presented here \cite{Trivedi2026_arXiv}. The mesh was further improved, specifically around the blades in the runner according to the recommendations provided by Choi and Moin \cite{Choi2012}. The final mesh selected for the LES simulations was 120 million nodes in the pump-turbine.

\begin{equation}
    {\mathrm{GCI}}_{\mathrm{fine}}^{21}
=
\frac{1.25\, e_{a}^{21}}{r_{21}^{k} - 1}.
\end{equation}

The validation of the numerical results was carried out with experimental data. The maximum measured efficiency of the turbine is $91.2\pm0.2\%$, $Q^\ast_{ED} = 1$ and $n^\ast_{ED} = 1$ \cite{Kerlefsen2025}. The validation error $(\hat{e}_v)$ is 5.6\%, where the experimental and numerical torque values are \SI{1049}{Nm} and \SI{1108}{Nm}, respectively. The validation error in pump mode operation is 9.1\%, where the experimental and numerical torque values are \SI{472}{Nm} and \SI{429}{Nm}, respectively.

For proper verification and validation, it is important to consider the total error $(\hat{e}_t)$ including the measurement uncertainties $(\hat{e}_{\mathrm{exp}})$, discretization error $(\hat{e}_{\mathrm{GCI}})$ and the validation error $(\hat{e}_v)$. The discretization (GCI) error is 3.5\%; the validation error is 5.6\%, and the experimental measurement error is 3.9\%. The experimental measurement error includes (1) uncertainties from the calibration of all instruments and sensors, (2) systematic and random errors, (3) repeatability errors. Detailed quantification of the experimental measurement error is presented in the literature \cite{Kerlefsen2025}. Thus, the total error in torque value obtained through numerical simulation is 7.7\% at the design load in turbine mode operation.

\begin{equation}
    \hat{e}_{v}
=
\frac{\lvert T_{\mathrm{num}} - T_{\mathrm{exp}} \rvert}{T_{\mathrm{exp}}} \times 100
\end{equation}

\begin{equation}
    \hat{e}_{t}
=
\sqrt{
    \hat{e}_{\mathrm{GCI}}^{2}
    +
    \hat{e}_{v}^{2}
    +
    \hat{e}_{\mathrm{exp}}^{2}
}
\end{equation}

\begin{table}[htbp]
\caption{Overview of the selected properties for computational modelling and the simulations for the present study.}
\label{tab:simulation_sequence}
\vspace{1em}
\centering
\resizebox{\textwidth}{!}{
\begin{tabular}{l l l l}
    \hline
    Parameter & Sequence 1 & Sequence 2 & Sequence 3 \\
    \hline
    Mesh size & 120 million & 120 million & 120 million \\
    Simulation type & Unsteady & Unsteady & Unsteady \\
    Turbulence model & SAS--SST & DES & LES WALE \\
    Wall function & Automatic & Automatic & Automatic \\
    Discretization scheme &
        High resolution ($0 < \beta < 1$) &
        Bounded Second Order Upwind ($\beta = 1$) &
        Bounded Central Difference \\
    Time step size &
        $0.5^\circ$ (\SI{3.193e-4}{\second}) &
        $0.1^\circ$ (\SI{6.386e-5}{\second}) &
        $0.01^\circ$ (\SI{6.386e-6}{\second}) \\
    Total time &
        10 revolutions &
        1 revolution &
        0.5 revolution \\
    Inner loop iterations & 5 & 5 & 10 \\
    Outer loop iterations & 7200 & 3600 & 18000 \\
    Maximum residual & $\le 10^{-4}$ & $\le 10^{-4}$ & $\le 10^{-4}$ \\
    \hline
\end{tabular}}
\end{table}

The verification and validation of the computational model are carried out using steady state results obtained from SST turbulence mode. The next phase of simulations is conducted using SAS-SST, DES, and LES WALE models. \autoref{tab:simulation_sequence} describes the proper sequence of simulations adopted for the present study. The unsteady simulations of SAS-SST are used to initialize the simulations of SAS-SST. The selected time-step size was 1 degree of runner rotation. The researchers used 0.5 – 2 degree time-step size depending on the requirement for the residual convergence \cite{Zobeiri2009,Binama2021,Xue2025,Xu2024,Cavazzini2016,Widmer2011,Xia2019}. The simulations carried out for a total of 10 revolutions of the runner. The next phase simulations were conducted using the DES approach. The simulations were initialized using results of SAS-SST. The simulations were conducted for one revolution of the runner with the time step size of \SI{0.1}{\degree}. The final simulations were conducted using LES WALE. The simulations were initialized using results of DES. Second order accurate scheme for space discretization and the second order backward Euler scheme for time discretization were used. The time step size was extremely small, i.e., \SI{0.01}{\degree} (\SI{6.386e6}{\second}), to ensure stability. Total time for the simulations was 0.5 revolution of the runner, which resulted in 18 000 iterations. The total combined inner and outer loop iterations were 180 000. The simulations with 120 million nodes required significant CPU time on the supercomputer.

\section{Results and discussions}
\label{sec:Results}
This study presents investigations on a reversible pump-turbine at speed-no-load conditions. We primarily focus on the inception of vortical flow, its transportation and interaction with the blades. At speed-no-load conditions, water flow rate through the turbine is small and the runner's angular speed is high. Consequently, massively separated flow in the blade passage. This section presents results and findings from LES WALE primarily and from DES where the time scale is larger than 0.5 revolution of the runner.

\subsection{Vortex inception and transport in turbine mode operation}
\label{subsec:Turbine}
\begin{figure}[htbp]
    \centering
    \includegraphics[width=1\linewidth]{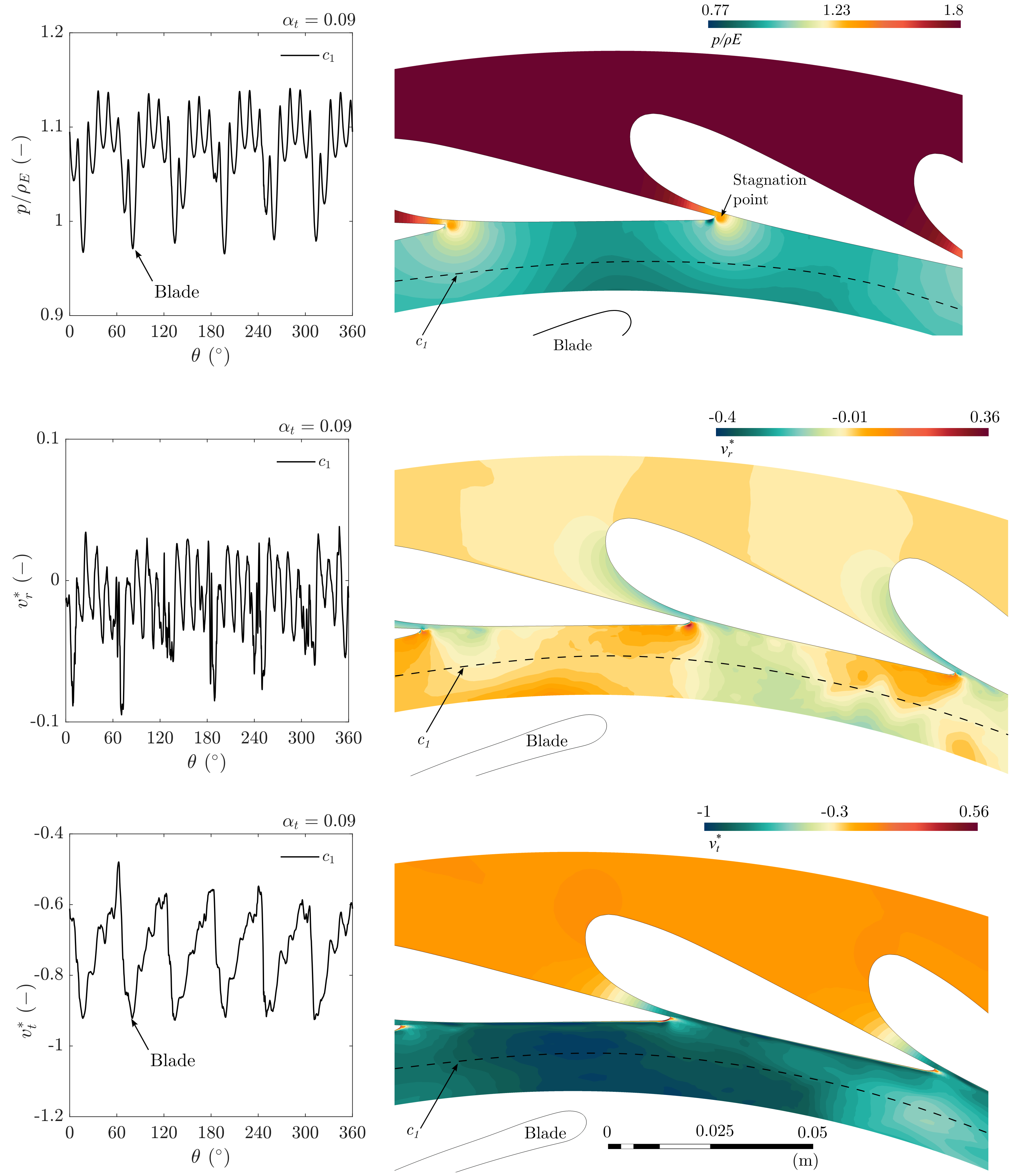}
    \caption{Static pressure and velocity (radial and tangential) variation at the runner inlet vaneless space. x-axis represents the 0 – 360 degrees of the runner circumference along the line $c_1$. Negative radial velocity indicates the vector towards the turbine axis. Negative tangential velocity indicates the vector towards the rotation speed direction of the runner.}
    \label{fig:Figure2}
\end{figure}

In turbine mode operation, the flow enters from the spiral casing, passes through stay vane and guide vane passages, then runner blades and leaves from the draft tube. The guide vane regulates the flow through the turbine, and the guide vane opening is 9\% of the design load for the present study. The 9\% opening is referred to as speed-no-load condition, where the blade produces sufficient mechanical torque to keep the turbine operating and to overcome the friction in the thrust and radial bearings. Static pressure difference across the guide vane passage is high, high pressure on the upstream side (spiral casing), and low pressure in the vaneless space (runner side). Furthermore, due to the small opening of the guide vanes, water flow through the guide vanes accelerates then decelerates quickly in the vaneless space. Interestingly, the vaneless space, the radial space between the guide vanes and the runner blades, is maximum at speed-no-load allowing water to decelerate fast — diffuser action — a perfect condition to form unsteady vortex. The available vaneless space is minimal at full load due to maximum opening of the guide vanes. \autoref{fig:Figure2} shows the variation of static pressure, radial velocity, and tangential velocity in the vaneless space. A polyline $(c_1)$ in the vaneless space along the runner circumference was created. The dimensionless radial $(r*)$ and axial $(z*)$ coordinates are 1.86 and 0, respectively. The polyline represents the \SI{360}{\degree} circumference of the runner at the final position. The pressure variation $(p/\rho E)$ clearly resembles the combined effect of the guide vane and the blade. The low pressure at $\theta = 17^\circ + \Delta 60^\circ$ indicates the blade leading edge. High pressure indicates the stagnation points of the guide vanes. The pressure changes periodically from one guide vane to another and reduces to minimum near the blade leading edge. Contour shows the combined pressure field in the guide vane passage and the vaneless space, where the stagnation point at the guide vane trailing edge is visible clearly, and low pressure at the blade leading edge location. Velocity, radial and tangential, variation is also presented in \autoref{fig:Figure2}. The velocity values are normalized characteristic velocity $(v_c)$. Peak-to-peak variations in radial velocity indicate the guide vane trailing edge, where the velocity is high. The variation is similar for all guide vanes. Contour shows high radial velocity at the trailing edge extending towards the vaneless space. The radial velocity in the small gap between two guide vanes is negative towards the runner and expands in the vaneless space. Near the blade leading edge, the velocity is positive and interacts with the guide vane trailing edge. Tangential velocity shows similar periodic variation in the vaneless space. However, the tangential velocity appears to be strongly dependent on the blade position and angular speed. Negative value indicates direction identical to the runner rotation. The low negative value indicates the blade position, it increases, and high in the middle of the passage. Unlike radial velocity, the impact of the guide vane trailing edge and the stagnation point is minimal on the tangential velocity.

\begin{equation}
    v_r^\ast = \frac{v_r}{\sqrt{2gH}}
\end{equation}

\begin{equation}
    v_t^\ast = \frac{v_t}{\sqrt{2gH}}
\end{equation}

\begin{figure}[htbp]
    \centering
    \includegraphics[width=1\linewidth]{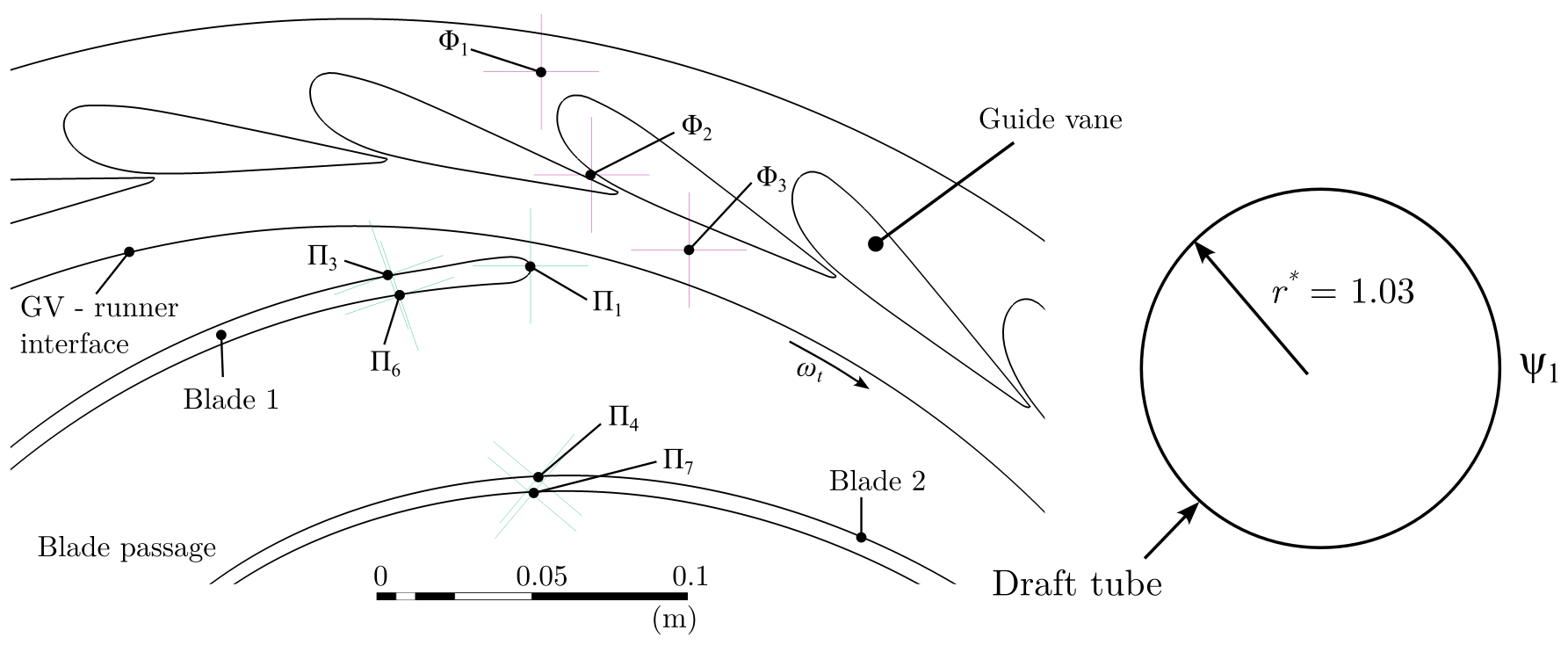}
    \caption{Visual representation and locations of observation points in the guide vane $(\Phi)$, runner $(\Pi)$ and draft tube $(\Psi)$.}
    \label{fig:Figure3}
\end{figure}

\begin{figure}[htbp]
    \centering
    \includegraphics[width=1\linewidth]{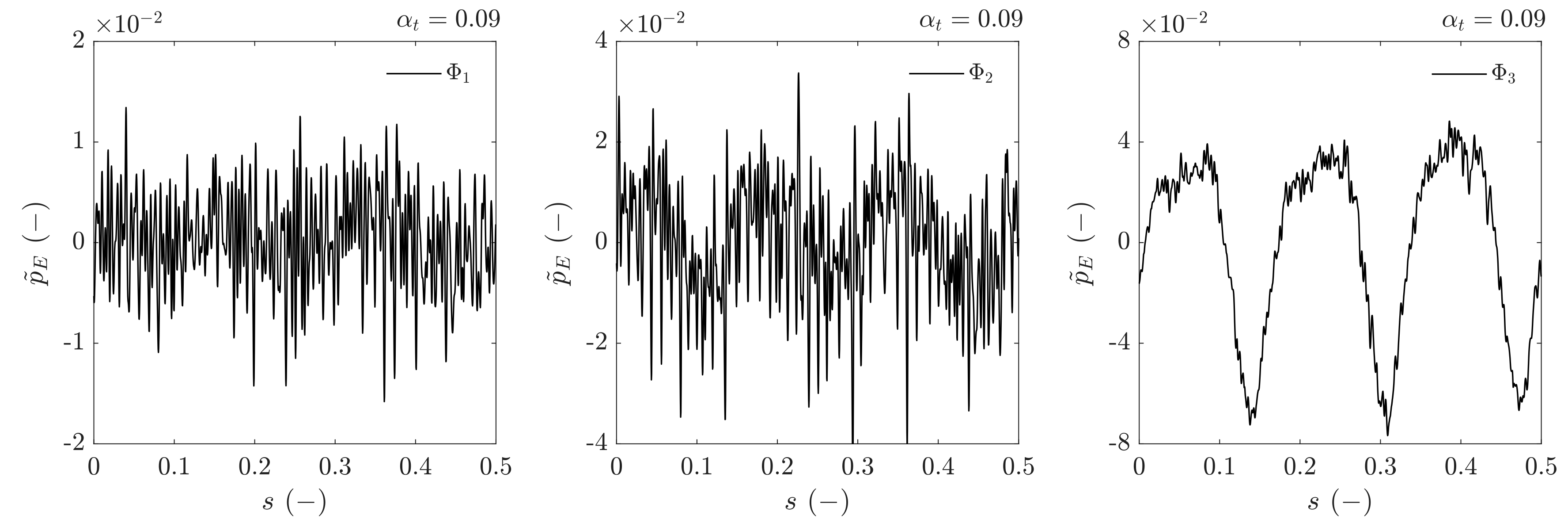}
    \caption{Unsteady pressure fluctuations in the guide vane passage. Scale on y-axis is adjusted for clear visualization of the fluctuations. Pitch $(s)$ on x-axis indicates the time-dependent runner rotation from \SI{0}{\degree} to \SI{180}{\degree}.}
    \label{fig:Figure4}
\end{figure}

To investigate time-dependent variation of flow characteristics, three observation points were created in the guide vane passage. The locations of the points are presented in \autoref{fig:Figure3}. Point $\Phi1$ is placed near the guide vane leading edge, a gap between the stay vane and the guide vane passage. Point $\Phi2$ is placed between the guide vanes. Point $\Phi3$ is placed in the vaneless space, following the trailing edge vortex shedding from the aligned guide vane. \autoref{fig:Figure4} shows time-dependent pressure fluctuations at $\Phi1$, $\Phi2$ and $\Phi3$ locations. Pressure fluctuations are normalized by the specific hydraulic energy $(\rho E)$ as shown in \autoref{eq:14}. The fluctuations were acquired when the simulation was running. Each sample corresponds to the time-step of the simulation, and the sampling frequency is 15.6 kHz. Acquired time vector data are converted into the runner angular rotation, referred to as pitch $(s)$; and $s = 1$ indicates one rotation of the runner. Hereafter, all time-dependent data are presented in terms of runner angular rotation for clarity and a meaningful correlation with an instantaneous blade position. The signature of pressure fluctuations at $\Phi1$, $\Phi2$ and $\Phi3$ locations is quite different. The fluctuations at $\Phi1$ location are predominantly wide band stochastic types. It is a combination of stay vane trailing edge vortex shedding and guide vane leading edge. Though a large part of the fluctuations at $\Phi2$ location are similar to the $\Phi1$ location, they are converging to the systematic frequency of runner blade passing $(f_b)$. The complete period of the blade passing frequency can be seen. Further downstream location, $\Phi3$, the fluctuations are clearly distinguishable and correspond to the blade passing frequency. The turbine includes 6 blade pump-turbine runner; we can see three complete periods for \SI{180}{\degree} runner rotations, $s = 0 - 0.5$. The other high frequency fluctuations are also obtained, which are related to the unsteady flow conditions in the vaneless space. At speed-no-load, the small opening of the guide vane, the trailing edge vortex is extended in the vaneless space. The location $\Phi3$ is aligned with the trailing edge of the preceding guide vane; therefore, the location $\Phi3$ acquires the effect of the trailing edge vortex. In addition, the location $\Phi3$ shows the combined effect of the interaction between the trailing edge vortex and the blade leading edge stagnation point.

\begin{equation}
\label{eq:14}
    {\tilde{p}}_E = \frac{p' - \bar{p}}{\rho E}
\end{equation}

\begin{equation}
    f_b = n \times Z_b
\end{equation}

\begin{figure}[htbp]
    \centering
    \includegraphics[width=1\linewidth]{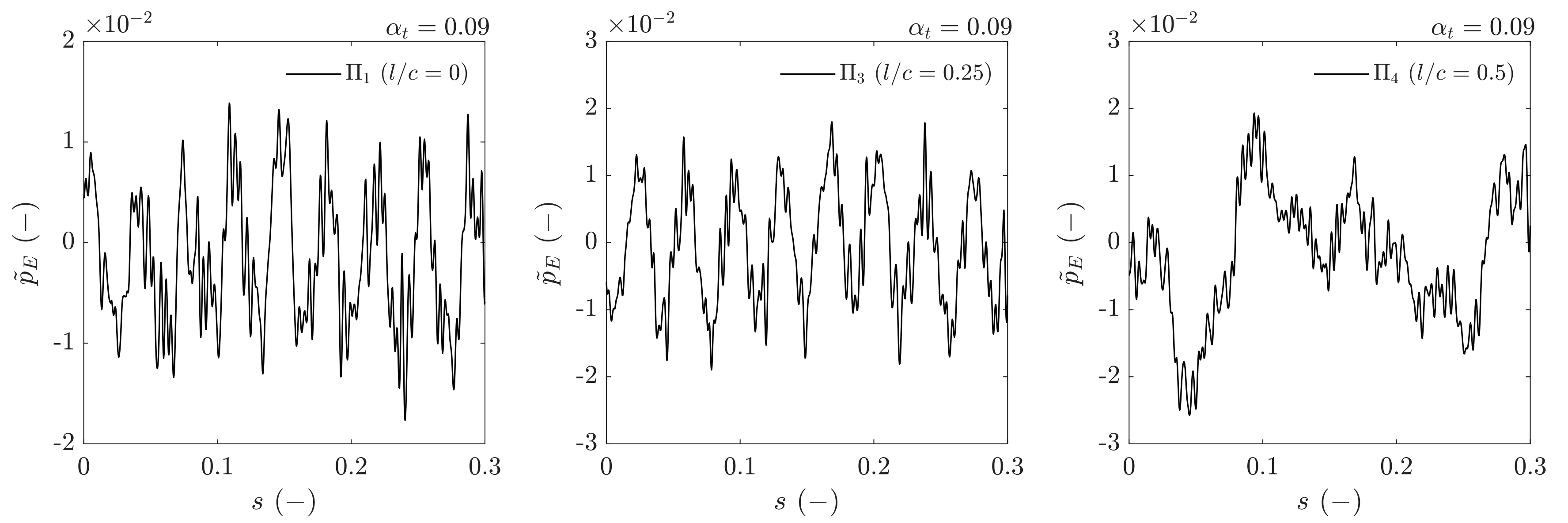}
    \caption{Unsteady pressure fluctuations in the runner blade passage. Scale on y-axis is adjusted for clear visualization of the fluctuations. Pitch $(s)$ on x-axis indicates the time-dependent runner rotation from \SI{0}{\degree} to \SI{108}{\degree}.}
    \label{fig:Figure5}
\end{figure}

In the stationary domains such as guide vanes and draft tube, we observe the blade passing frequency and its harmonic due to rotor-stator interaction. In the rotation domain, such as runner, we observe the guide vane passing frequency and its harmonic. To investigate the unsteady flow in the runner domain, 8 observation points were created at mid-span $(\lambda = 0.5)$ of a blade. Two points, $\Pi1$ and $\Pi2$, are located at the blade leading edge and trailing edge, respectively. Three points, $\Pi3$, $\Pi4$, and $\Pi5$, are located on the blade high-pressure side at a chord length $(l/c)$ of 0.25, 0.5, and 0.75, respectively. Three points, $\Pi6$, $\Pi7$, and $\Pi8$ are located on the blade low-pressure side at a chord length $(l/c)$ of 0.25, 0.5, and 0.75, respectively. Unsteady time-dependent pressure fluctuations were investigated at all these observation points. \autoref{fig:Figure5} shows the pressure fluctuations at the selected three locations, $\Pi1$, $\Pi3$ and $\Pi4$. The fluctuations are shown for $s = 0.3$, i.e., \SI{108}{\degree} runner blade rotations. The location $\Pi1$ is, at the leading edge of the blade, showing the pressure fluctuations of the guide vane passing frequency $(f_{gv})$ of 121.6 Hz and the first harmonic, 243.7 Hz. In addition, the other high frequency, wide band, fluctuations are acquired, indicating the stochastic nature of flow and separation from the leading edge. At further downstream location, $\Pi3$, the fluctuations are of a similar signature. The large variation in the fluctuations is obtained at $\Pi4$ location and towards the trailing edge. The guide vane passing frequency is no longer predominant. The fluctuations are clearly stochastic, both low- and high-frequency, revealing the presence of a highly unsteady vortical flow.

\begin{figure}[htbp]
    \centering
    \includegraphics[width=1\linewidth]{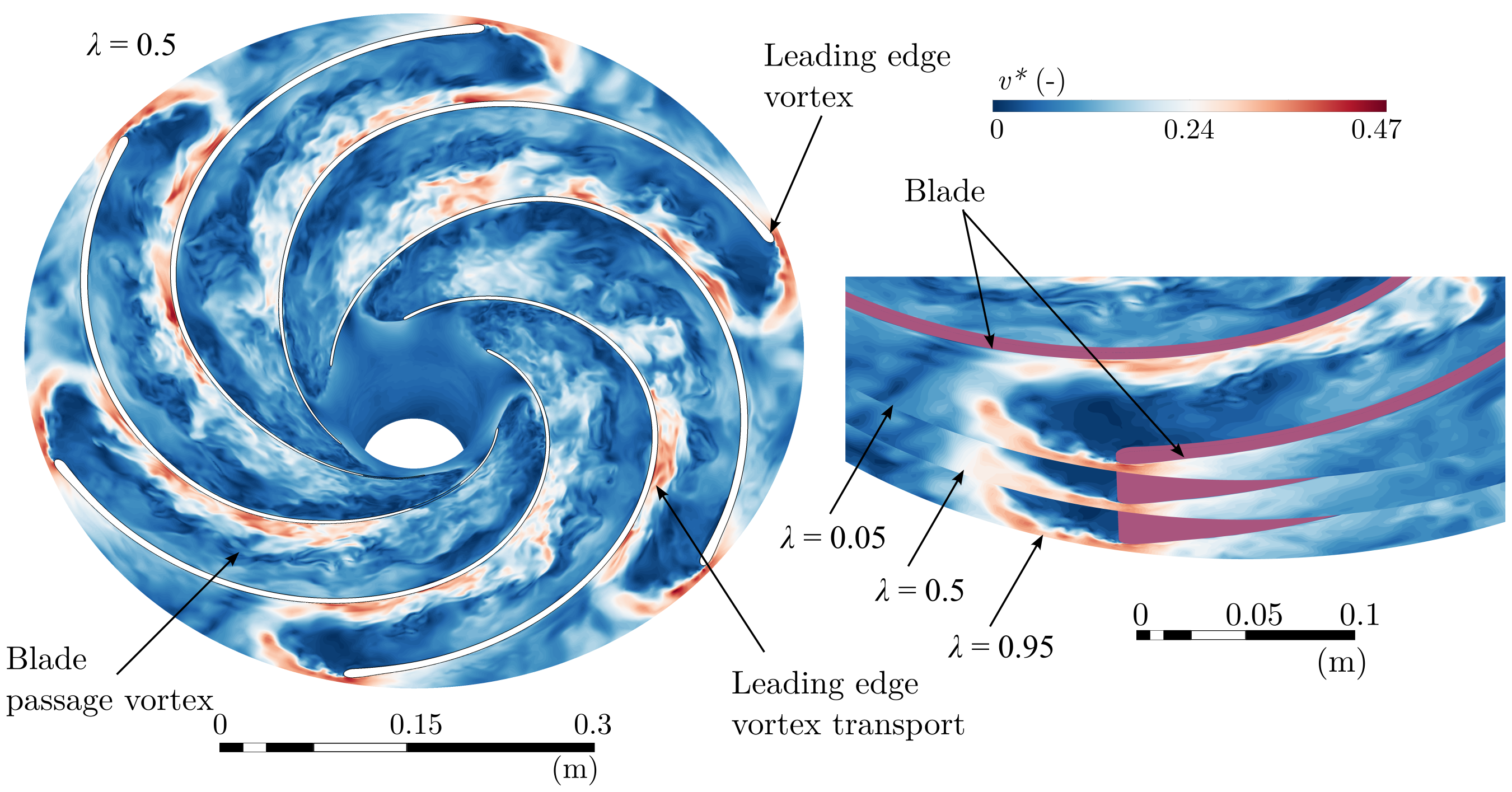}
    \caption{Contours of dimensionless velocity showing inception and transport of leading-edge vortex in the blade passage.}
    \label{fig:Figure6}
\end{figure}

\begin{equation}
    f_{gv} = n \times Z_{gv}
\end{equation}

\begin{equation}
\label{eq:17}
    v^\ast = \frac{\sqrt{u^2 + v^2 + w^2}}{\sqrt{2gH}}
\end{equation}

\begin{figure}[htbp]
    \centering
    \includegraphics[width=1\linewidth]{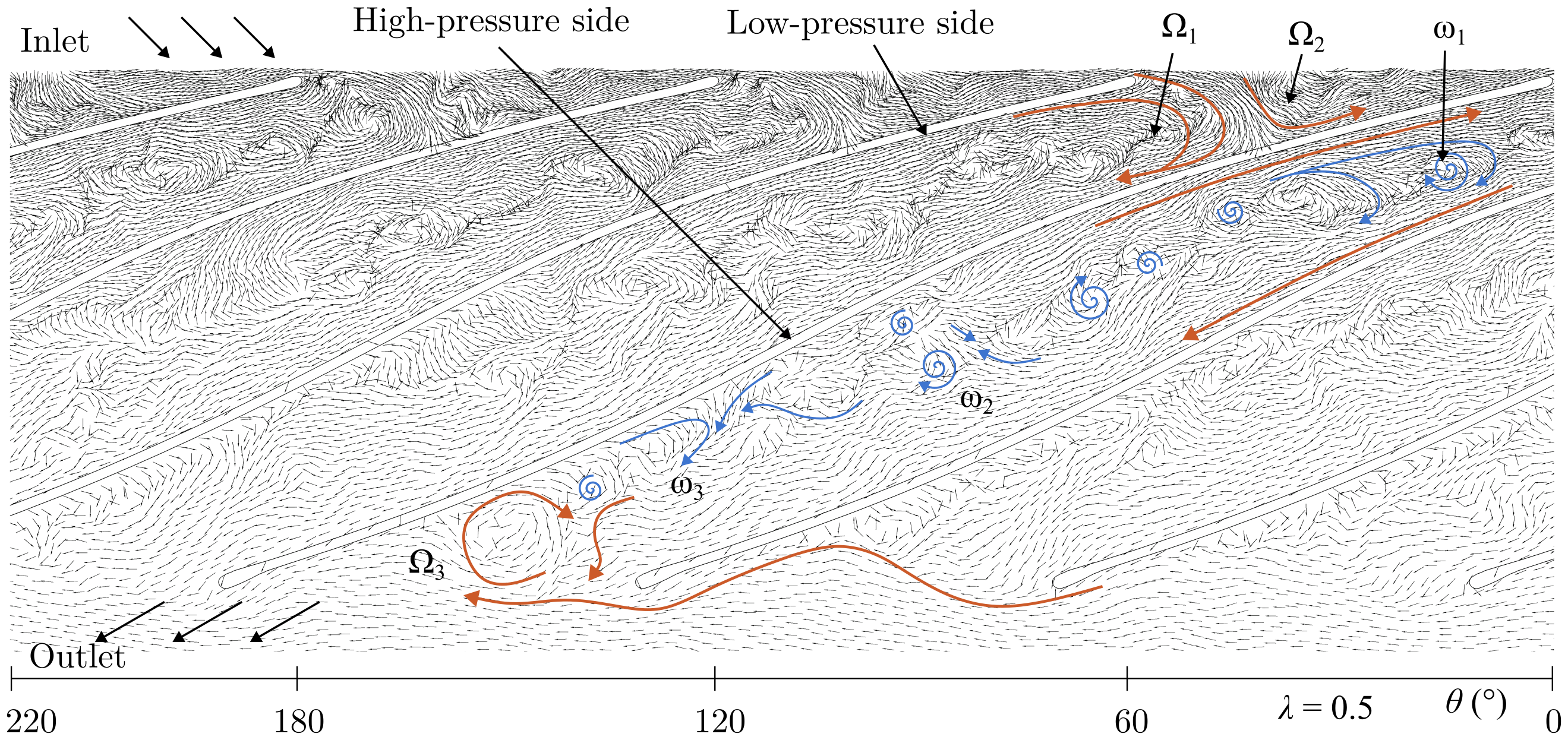}
    \caption{Vortex transport and interaction with the blades. $\Omega$ represents the bulk flow vortex transport; and $\omega$ represents the local flow vortex transport within blade passage.}
    \label{fig:Figure7}
\end{figure}

To investigate the unsteady vortex transport and the interaction with the blades, contours representing the flow field in the blade passage are obtained. \autoref{fig:Figure6} shows the contours of dimensionless velocity at mid-span $(\lambda = 0.5)$ of the blades. The contours are normalized with a characteristic velocity; see \autoref{eq:17}. Water from the guide vane passage leaves at a very low angle and enters the blade passage. The small angle of incoming flow initiates leading edge separation, sets on a large longitudinal vortex along the blade's leading edge. The longitudinal vortex transports towards the high-pressure side of the blade and splits into two vortical zones, marked as $\Omega1$ and $\Omega2$. The longitudinal vortex is extended over the entire span of the blade, which is evidently presented as contour slices at $\lambda = 0.05$, $\lambda = 0.5$ and $\lambda = 0.95$. The span values $\lambda = 0.05$ and $\lambda = 0.95$ indicate 5\% span from the hub wall and 5\% span from the shroud wall. The massive longitudinal vortex breaks down and travels downstream into the blade passage, which can be seen as a region with high velocity along the high-pressure side of the blade, marked as ``Leading edge vortex transport". At same time, a ``string of swirls" is developed in the blade passage, predominantly along the low-pressure side of the blade. To characterize the vortex pattern and its transport phenomenon, velocity vectors at $\lambda = 0.5$ are acquired and presented in \autoref{fig:Figure7}. It is important to note that the presented vortex pattern corresponds to the instantaneous time stamp—last time step of the simulation, and the relative position of the blades with respect to the stationary guide vanes. The signature of the vortex pattern changes with the blade position. Thought, the vectors are shown for $\lambda = 0.5$, no significant deviation is observed longitudinally other than change of spatial locations of the ``string of swirls". The large vortex pattern is marked with `$\Omega$', and the string of swirls is marked with `$\omega$'. The main longitudinal vortex $(\Omega1)$ travels from the leading edge towards the blade passage and strikes on the high-pressure side of the neighboring blade. Consequently, it splits and develops another vortex, marked as `$\Omega2$', and transports upstream towards the neighboring blade’s leading edge. Flow attached to the high-pressure side of the blade is marching forward, whereas the flow attached to the low-pressure side of the blade is reversible. This contra‑directional flow induces an adverse pressure gradient on the low‑pressure side and a favorable pressure gradient on the high‑pressure side, thereby triggering the inception of a swirl‑string (vortex) structure. The vortex $\omega1$ is the secondary swirl from the vortex $\Omega1$, and spins in a clockwise direction locally. We can see other small vortices along the blade passage, and the majority of them spin locally in a clockwise direction. At further downstream in the blade channel (around $l/c = 0.7$), two parallel vortices appear, forming a pair of twin vortices, marked as $\omega2$. Interestingly, flow on high-pressure side and low-pressure side merge at this location. This results in the collapse of organized vortices and the inception of a highly stochastic and reversible flow, marked at $\omega3$. Finally, at the outlet of the blade passage, a large vortical region $\Omega3$ is formed. The flow from the neighboring passage is pumped into this passage. This results in the formation of a large vortical region near the low-pressure side of the blade and creates a large blockade at the outlet. The region $\Omega3$ spins in clockwise direction and interacts with the upcoming stochastic flow of twin vortices, $\omega3$. In addition, the strong reversible (pumping) flow towards the runner from the draft tube was obtained. The region $\Omega3$ was not obtained in all blade passages. This suggests that the region $\Omega3$ appears time-dependent and dependent on the instantaneous spatial position of the runner relative to the guide vanes. Further analysis of the results is conducted in the draft tube focusing on the characteristics of the reversible (pumping) flow.

\begin{figure}[htbp]
    \centering
    \includegraphics[width=1\linewidth]{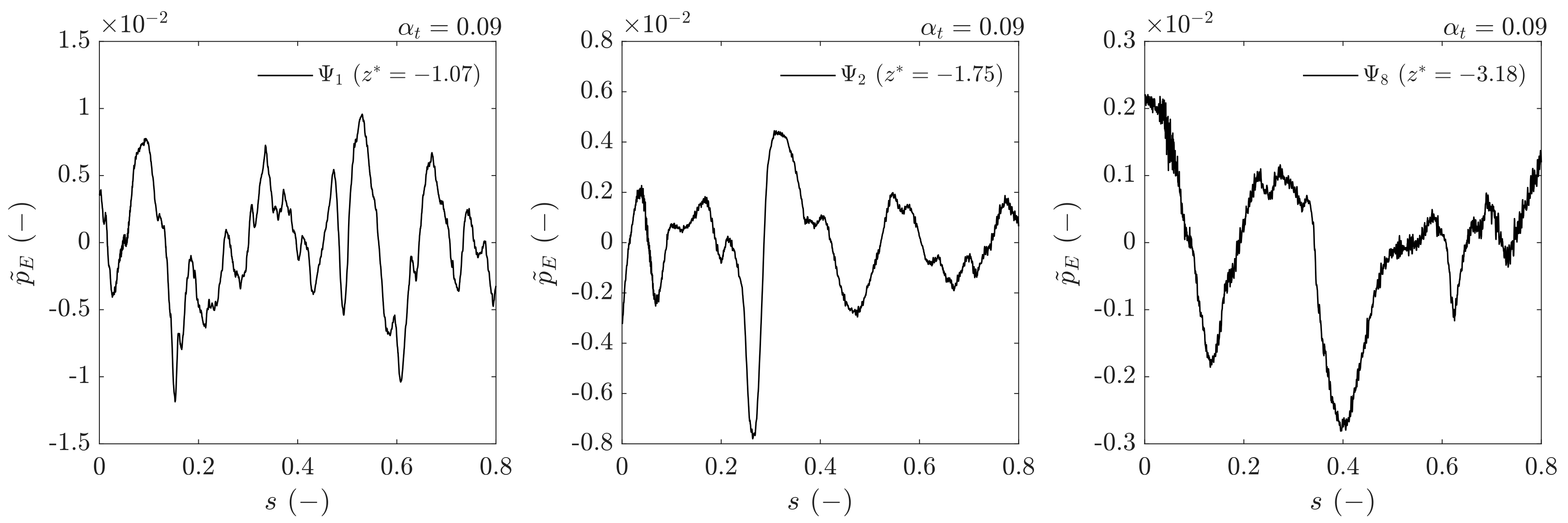}
    \caption{Unsteady pressure fluctuations in the draft tube. Pitch $(s)$ on x-axis indicates the time-dependent runner rotation from \SI{0}{\degree} to \SI{288}{\degree}. Scale on y-axis is adjusted for clear visualization of the fluctuations.}
    \label{fig:Figure8}
\end{figure}

Draft tube is generally designed to recover pressure energy from the available kinetic energy and help to minimize cavitation. Therefore, the draft tube is designed such that its cross-sectional area increases (diffuser) along the flow path. Flow at design conditions is stable in the draft tube; however, at speed-no-load, adverse pressure gradient is predominant. Moreover, tangential velocity at the runner outlet is high, which provides angular momentum to the near wall flow of the draft tube. Pressure fluctuations at three locations are presented in \autoref{fig:Figure8}. The fluctuations correspond to the locations $\Psi1$, $\Psi2$ and $\Psi3$, and their respective three levels $(z/R)$ in the draft tube are $z* = -1.07$, $z* = -1.75$ and $z* = -3.18$. The location $\Psi1$ is at the interface connecting the runner and the draft tube; thus, pressure fluctuations correspond to the runner outlet flow field and the blade passing frequency $(f_b)$. Systematic fluctuations of $f_b$ are not acquired clearly as the blade trailing edge is somewhat away from the observation location. The fluctuations also indicate the strong presence of secondary flow. The locations $\Psi2$ and $\Psi3$ show the minimal effect of the blade with small amplitude variation of the blade passing frequency. In addition, $\Psi2$ and $\Psi3$ show high frequency wide band fluctuations indicating the presence of secondary flow of stochastic characteristics.

\begin{figure}[htbp]
    \centering
    \includegraphics[width=1\linewidth]{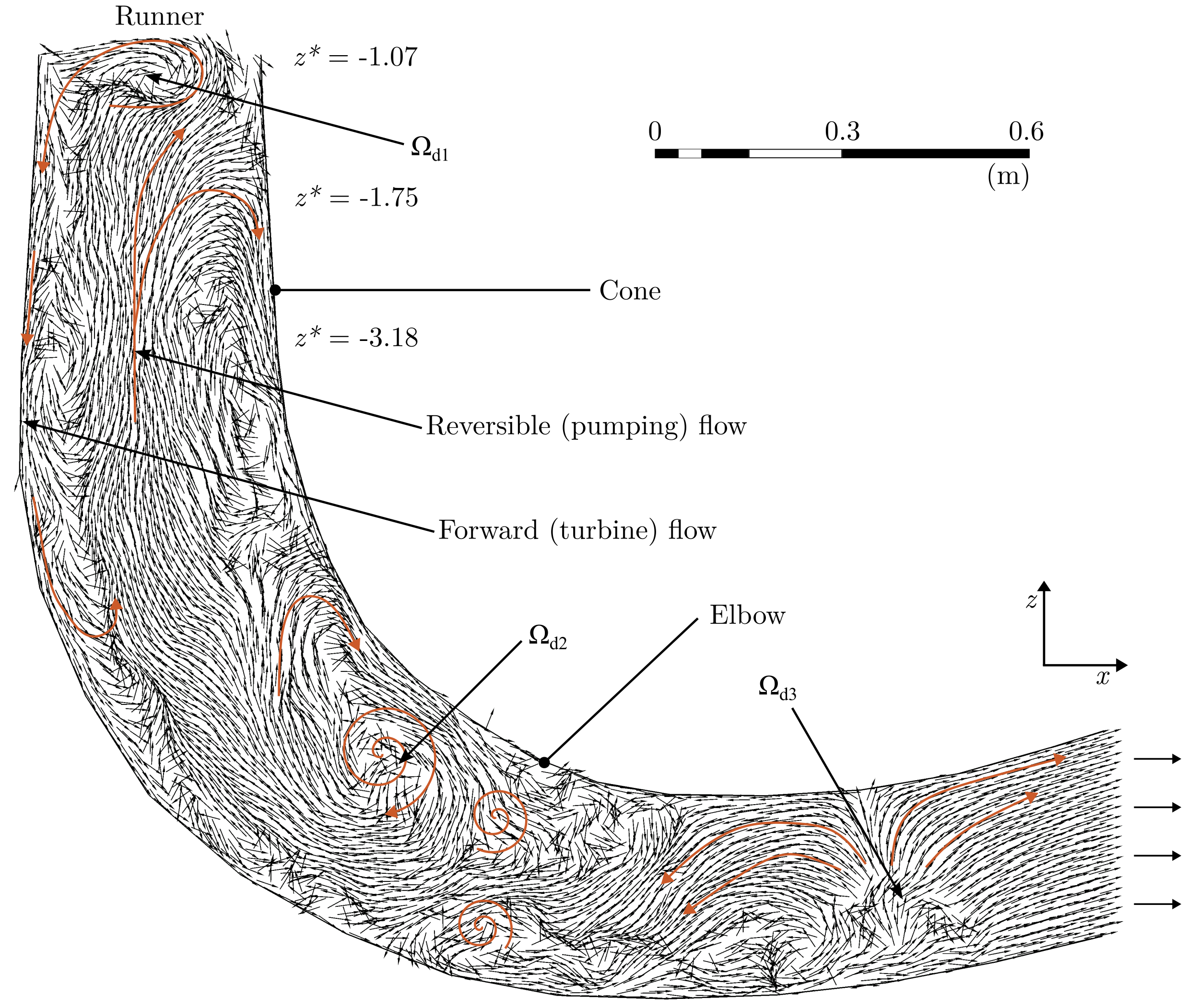}
    \caption{Vortex transport in the draft tube. $\Omega_d$ represents the bulk flow vortex transport. Outlet of the draft tube is shown to be short due to limited space in x-direction.}
    \label{fig:Figure9}
\end{figure}

\begin{figure}[htbp]
    \centering
    \includegraphics[width=1\linewidth]{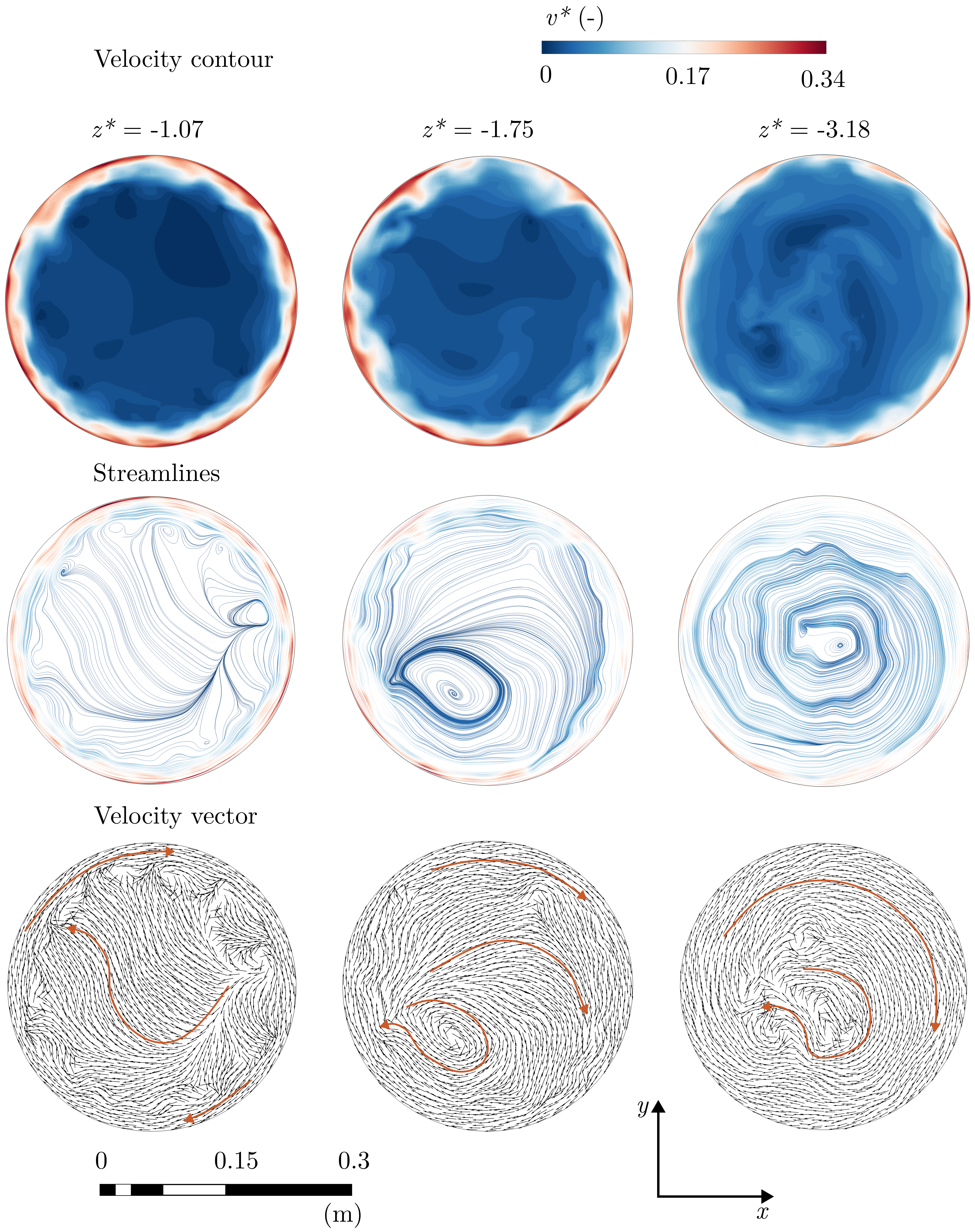}
    \caption{Velocity contours, streamlines, and vectors in draft tube radial plane at $z* = -1.07$, $z* = -1.75$, and $z* = -3.18$. The center of the plane follows the turbine rotational axis, and it is located at the point (x, y) = (0, 0).}
    \label{fig:Figure10}
\end{figure}

Velocity vectors on the xz-plane of the draft tube are acquired to study characteristics of the stochastic secondary flow in the axial direction of flow. \autoref{fig:Figure9} shows the acquired vectors resembling the secondary flow in the draft tube. Interestingly, the main flow in the draft tube core section is reversible (pumping) and flowing towards the runner. Source of the reversible flow appears at the elbow of the draft tube, marked as vortex $\Omega_{d3}$. Flow from the vortex $\Omega_{d3}$ divides into two directions: towards runner — reversible — and towards outlet. Several swirling zones of secondary flow exhibit in the elbow. Another large vortex $\Omega_{d2}$ near small radii of the bend can be seen besides the swirling zones, rotating in a clockwise direction. Wall-attached flow on small radii of the bend is approaching downward while the main flow in the core (center) of the draft tube is reversible. It appears that the contra-directional flow on two sides provides required momentum to the vortex $\Omega_{d2}$. Furthermore, several eddies of secondary flow appear on the outer radii of the elbow and those are extended up to the runner outlet. Interestingly, flow direction on the outer radii wall is traveling forward (turbine). Thus, two opposite directions of flow induce local separation and vortical zones in the axial direction of the draft tube. We can see continuous mixing and separation of the core reversible flow and the forward flow. On the interface at the runner outlet and draft tube inlet, a large vortical zone $\Omega_{d1}$ is obtained. The rotational direction is counterclockwise of the zone $\Omega_{d1}$. When the reversible flow reaches the runner-draft tube interface, it creates a vortical zone as flow from the runner blade passage is forward and transports downward. Flow on the outer edge of the runner is forward with very high tangential velocity, which is moving downward. The combined effect results in the formation of the large vortical region $\Omega_{d1}$ in the center of the draft tube and blocking the main incoming flow from the runner. To investigate the characteristics of secondary flow and the vortical regions, xy-planes are created at three axial locations, $z* = -1.07$, $z* = -1.75$, and $z* = -3.18$. \autoref{fig:Figure10} shows the acquired velocity contours, streamlines, and vectors at these locations. We can see high velocity near the draft tube wall at all locations. The plane $z* = -1.07$ shows the effect of the blade trailing edge vortex shedding. High rotational speed of the runner and the predominant effect of the centrifugal force drive the bulk flow towards the outer wall, shroud, and draft tube wall. At the same time, high tangential velocity appears exchanging momentum downward from the runner to the draft tube. The streamlines and vector clearly show the presence of a vortical zone on the xy-plane. On $z* = -1.07$, several swirling zones of small vortex appear, where the main flow from the runner outlet rotates in a clockwise direction. i.e., similar to the runner angular rotation direction. The reversible flow in the center of the plane is redirected towards the wall. As a result, the interaction between these two opposite flow forms separating swirling zones. On $z* = -1.75$, we can see that the streamlines and vectors are spiraling around a large vortical zone. On $z* = -3.18$, the vortical zone is shifted to the center of the plane, and the bulk flow is swirling around it, including the wall-attached flow. It appears that the vortical region remains in center for the rest of the axial locations $(z* > -3.18)$ towards the elbow, and the bulk flow rotates in clockwise direction around this vortical zone.

\subsection{Vortex inception and transport in pump mode operation}
\label{subsec:Pump}
The direction of water flow and turbine runner rotation are reversed during pump mode operation of a reversible pump-turbine. The draft tube is where water enters, it passes through the runner, then through the guide vane and stay vane passages, and finally exits the spiral casing. Flow direction in pump mode is shown in \autoref{fig:Figure1}. The flow field is different from that of the turbine mode, e.g., blade leading edge vortex shedding towards the vaneless space, vortex shedding from the guide vane trailing edge towards the stay vane passage. Analyzed results of pump mode simulations are presented in logical order of water flow direction. We first present the main findings from the draft tube, followed by those from the runner blade passage and its interaction with the unsteady vortex transport, and finally the results from the guide vane and spiral casing. In this section, the naming conventions for the observation points, leading edge, trailing edge, chord length, spatial locations, and other reference dimensions follow the naming of the turbine mode operation for clarity and understanding of the flow field.

\begin{figure}[htbp]
    \centering
    \includegraphics[width=1\linewidth]{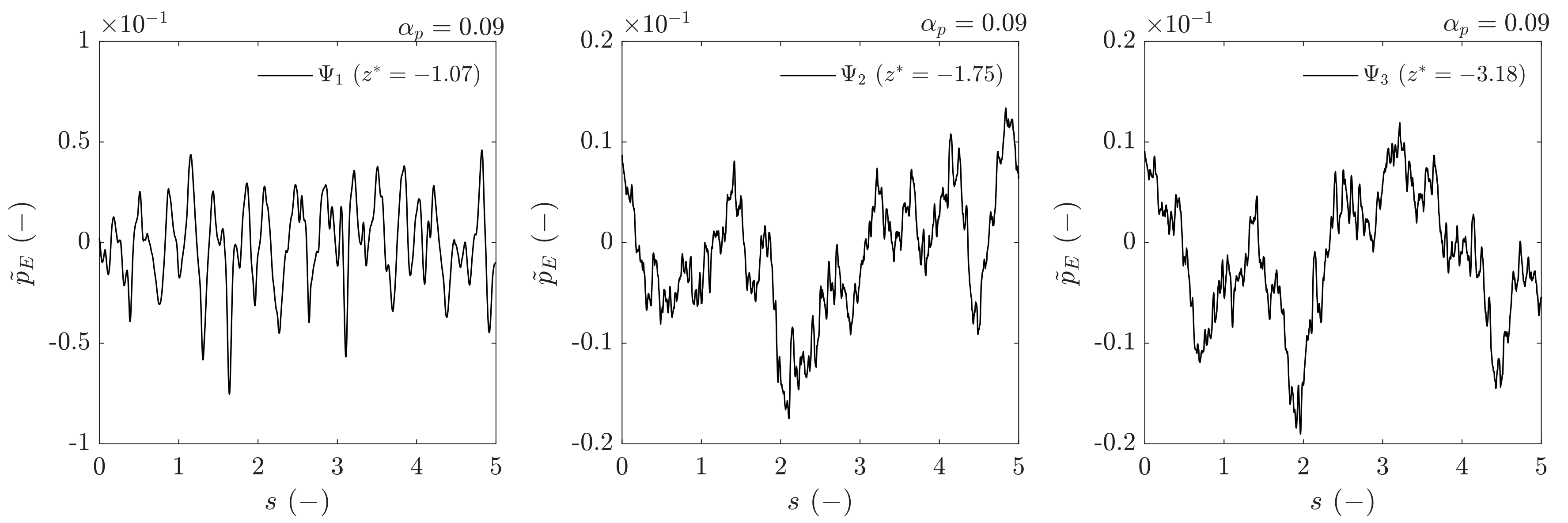}
    \caption{Unsteady pressure fluctuations in the draft tube during five rotations of the runner. Pitch $(s)$ on x-axis indicates the time-dependent runner rotation. Scale on y-axis is adjusted for clear visualization of the fluctuations.}
    \label{fig:Figure11}
\end{figure}

The investigated reversible pump-turbine is equipped with an elbow-type draft tube. Three observation points are created at different axial locations in the draft tube. \autoref{fig:Figure11} shows unsteady pressure fluctuations in the draft tube. Point $\Psi1$ is placed near the runner edge joining the draft tube $(z* = -1.07)$. The point shows largely systematic pressure fluctuations pertaining to the runner angular speed $(f_n)$ and its harmonics. Points $\Psi2$ and $\Psi3$ are placed further downstream between the runner edge and the elbow at the axial distance of $z* = -1.75$ and $z* = -3.18$, respectively. Both points show stochastic type fluctuations with no specific signature suggesting predominantly secondary flow. This is associated with the effect of elbow and separation from the small radii of the elbow. Both points are located on the side, which is axially aligned with the small radii of the elbow. They acquire fluctuations related to the separation from the elbow in pump mode operation.

\begin{figure}[htbp]
    \centering
    \includegraphics[width=1\linewidth]{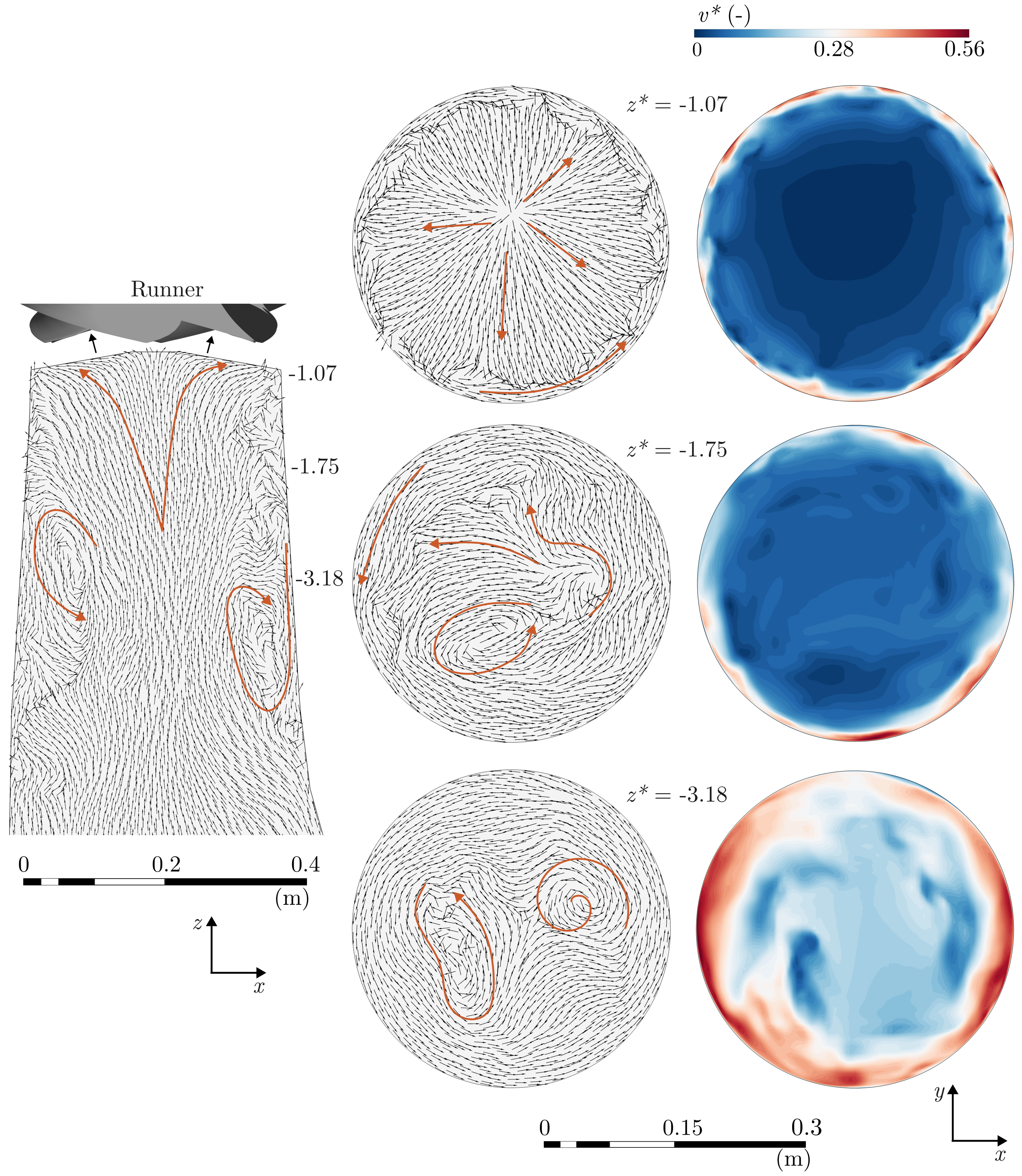}
    \caption{Vectors and dimensionless velocity in the draft tube cone section showing the secondary flow towards the runner.}
    \label{fig:Figure12}
\end{figure}

The analyzed flow field in the draft tube is presented in \autoref{fig:Figure12}. Dimensionless velocity and the velocity vectors in the draft tube cone $(z* = -1.07$ to $z* = -6.1)$ are shown. The velocity is normalized using \autoref{eq:17}. Contours of velocity and vectors are shown for three axial planes representing the locations $\Psi1$, $\Psi2$ and $\Psi3$ of \autoref{fig:Figure11}. The core flow in the center of the draft tube flows towards the runner; however, near wall flow instability is observed and it transports opposite to the core flow. This develops a separation line between the core flow and near wall flow, and causes wall attached local swirls. Near the runner edge, $z* = -1.07$, core flow transports outward, whereas the near wall flow rotates in a counterclockwise direction, following the runner. At the intermediate plane, $z* = -1.75$, flow instability is increased, where a large vortical region is established rotating in a counterclockwise direction. The core flow is divided into two zones and redirected towards the wall. Furthermore, the region of near wall is increased as compared to the plane, $z* = -1.07$. At further downstream, $z* = -3.18$, a pair of vortices is established, which divides the main flow in the core region. Both vortices spin in the opposite direction, and the vortices gain momentum from near wall flow.

\begin{figure}[htbp]
    \centering
    \includegraphics[width=0.8\linewidth]{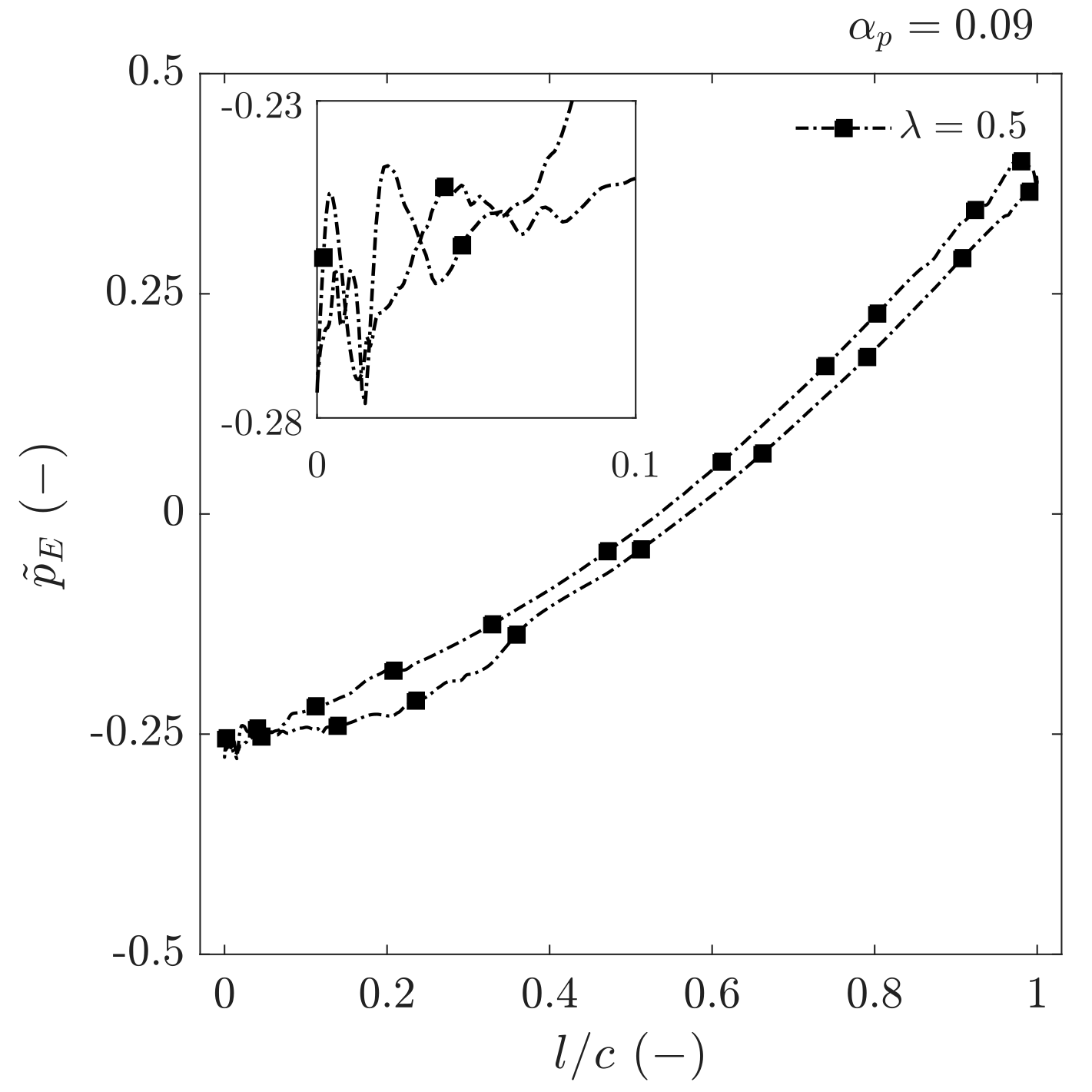}
    \caption{Blade loading at mid-span of the blade. Chord length $l/c = 0$ indicates blade trailing edge.}
    \label{fig:Figure13}
\end{figure}

\begin{figure}[htbp]
    \centering
    \includegraphics[width=1\linewidth]{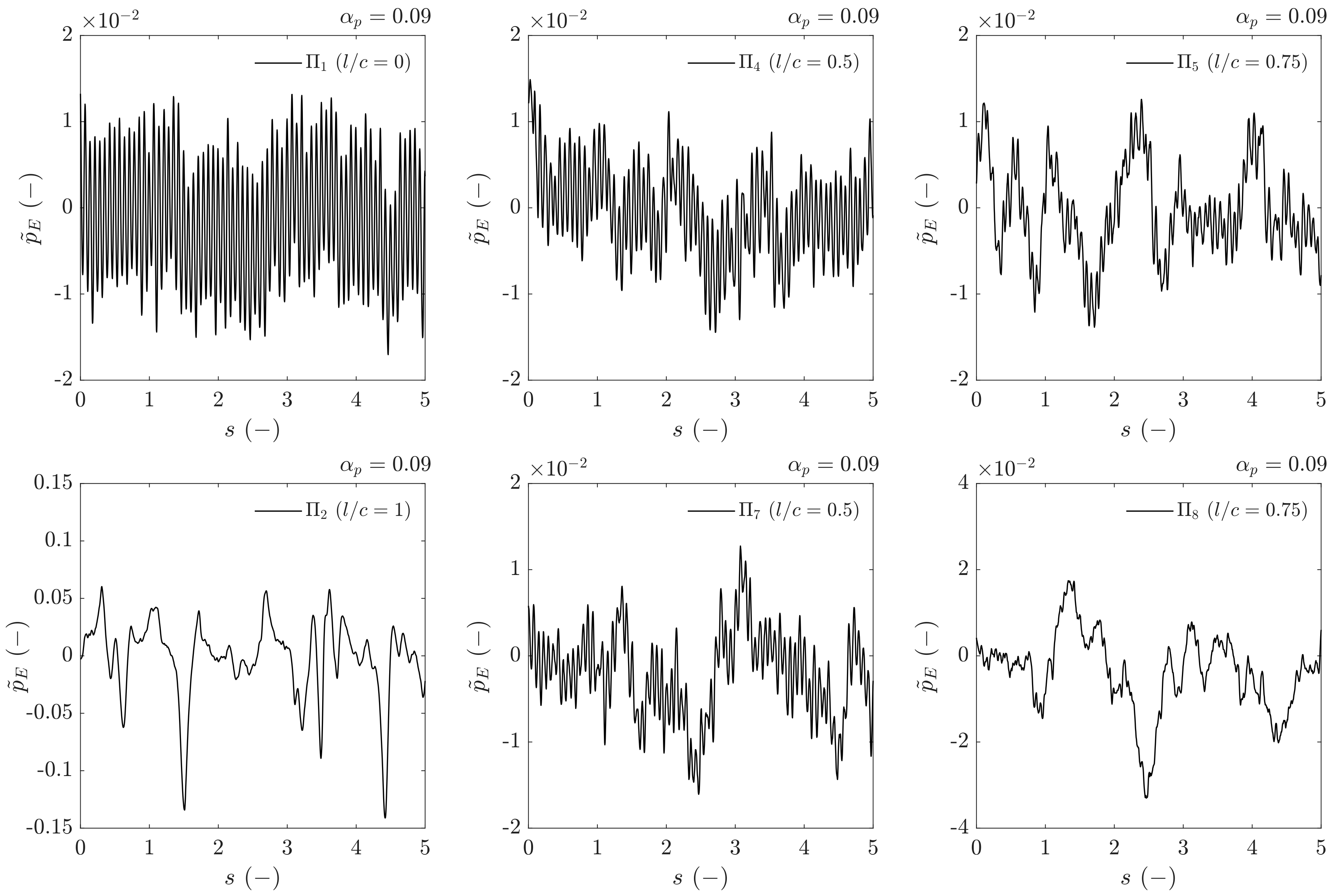}
    \caption{Unsteady pressure fluctuations on the high- and low-pressure sides of the blade during five rotations of the runner. Pitch $(s)$ on x-axis indicates the time-dependent runner rotation. Scale on y-axis is adjusted for clear visualization of the fluctuations.}
    \label{fig:Figure14}
\end{figure}

Flow travels from the draft tube to the runner and passes through the blade passage trailing edge and leaves from the leading edge side. At speed-no-load, the head developed by the pump-turbine is minimal. Flow instabilities in the blade passage are extreme due to strong separation of flow. At the same time, a small pressure difference exists between the high- and low-pressure sides of the blade. \autoref{fig:Figure13} shows the blade loading at mid-span $(\lambda = 0)$ of the blade at speed-no-load condition. Chord length $(l/c) = 0$ represents the trailing edge of the blade. As we can see, the loading increases from the trailing edge to the leading edge. However, the difference between the high- and low-pressure sides is small. On the trailing edge, the loading remains asymmetric, undergoing continuous alternation up to $l/c = 0.1$. The alternate loading on the trailing edge is shown in the scaled plot within \autoref{fig:Figure13}. The blade loading corresponds to the results obtained from the last time-step of the simulation. To investigate the loading over several time-steps, we created observation points on both sides of the blade at selected locations; the locations are shown in \autoref{fig:Figure3}. The time-dependent pressure fluctuations on the high- and low-pressure sides are shown in \autoref{fig:Figure14}. The observation points $\Pi1$ and $\Pi2$ are placed on the blade leading edge and trailing edge, respectively. Fluctuations at $\Pi1$ location are mainly systematic and imply the guide vane passing frequency $(f_{gv})$ and its higher harmonics. The fluctuations at $\Pi2$ location are stochastic and no specific signature is obtained. We can see a large drop in pressure periodically at a frequency of around 7 Hz, which appears to be the runner angular speed (6.8 Hz). It is difficult to locate the exact frequency due to the short length of the acquired pressure data from the simulations as compared to the sampling rate of 24 kHz. Observation points $\Pi4$ and $\Pi5$ are placed on the high-pressure side at the chord lengths 0.5 and 0.75 at mid-span, respectively. Fluctuations at $\Pi4$ are largely systematic, similar to that of the $\Pi1$ location; in addition, we can see other low frequency variations of bulk frequencies, potentially associated with unsteady flow in the blade passage. Fluctuations at $\Pi5$ location are both systematic with high frequency and stochastic with low frequency. When comparing the low-pressure side locations, $\Pi7$ and $\Pi8$, opposite to the $\Pi4$ and $\Pi5$ locations, the fluctuations are mainly stochastic, both high and low frequencies suggesting predominantly unsteady secondary flow.

\begin{figure}[htbp]
    \centering
    \includegraphics[width=1\linewidth]{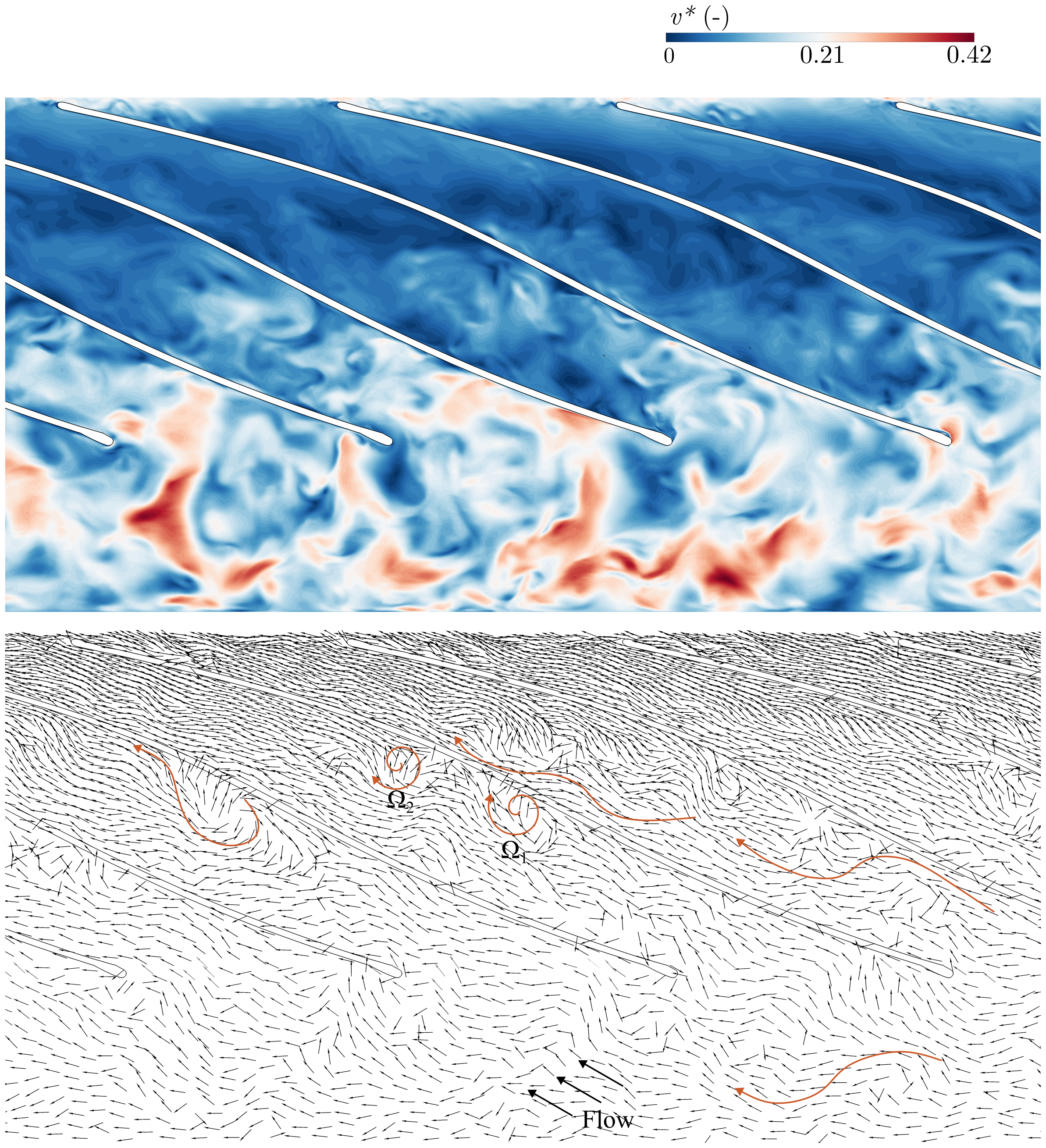}
    \caption{Vectors and dimensionless velocity in the blade passage showing unsteady vortex transport from blade trailing edge to the leading edge side.}
    \label{fig:Figure15}
\end{figure}

In a pump mode, flow enters the runner from the trailing edge side, where flow separation is high due to adverse pressure gradient set by the blade curvature. To study flow separation and the vortex transport in the blade passage, contours of velocity and the vectors are acquired from the simulation results, which are presented in \autoref{fig:Figure15}. The contours are extracted at mid-span of the blade, and blade-to-blade view of the runner is presented. Strong flow instabilities are obtained around the trailing edge of the blades. A series of unsteady vortices is developed soon after the flow enters the blade passage. Two large vortical zones can be seen, those are marked with $\Omega1$ and $\Omega2$, and both are on the low-pressure side of the blade. Furthermore, flow around these zones is reversible and transports towards the trailing edge and creates partial blockades. The vortical zones are not persistent and they are temporal, where the instantaneous location is dependent on the instantaneous position of the blade passage relative to the guide vane. The flow travels further upstream to the outlet of the passage $l/c = 0.6 - 1$. The pressure and velocity field is relatively stable and largely driven by the rotor-stator interactions, as evident from the acquired data at $\Pi1$ and $\Pi4$ locations.

\begin{figure}[htbp]
    \centering
    \includegraphics[width=1\linewidth]{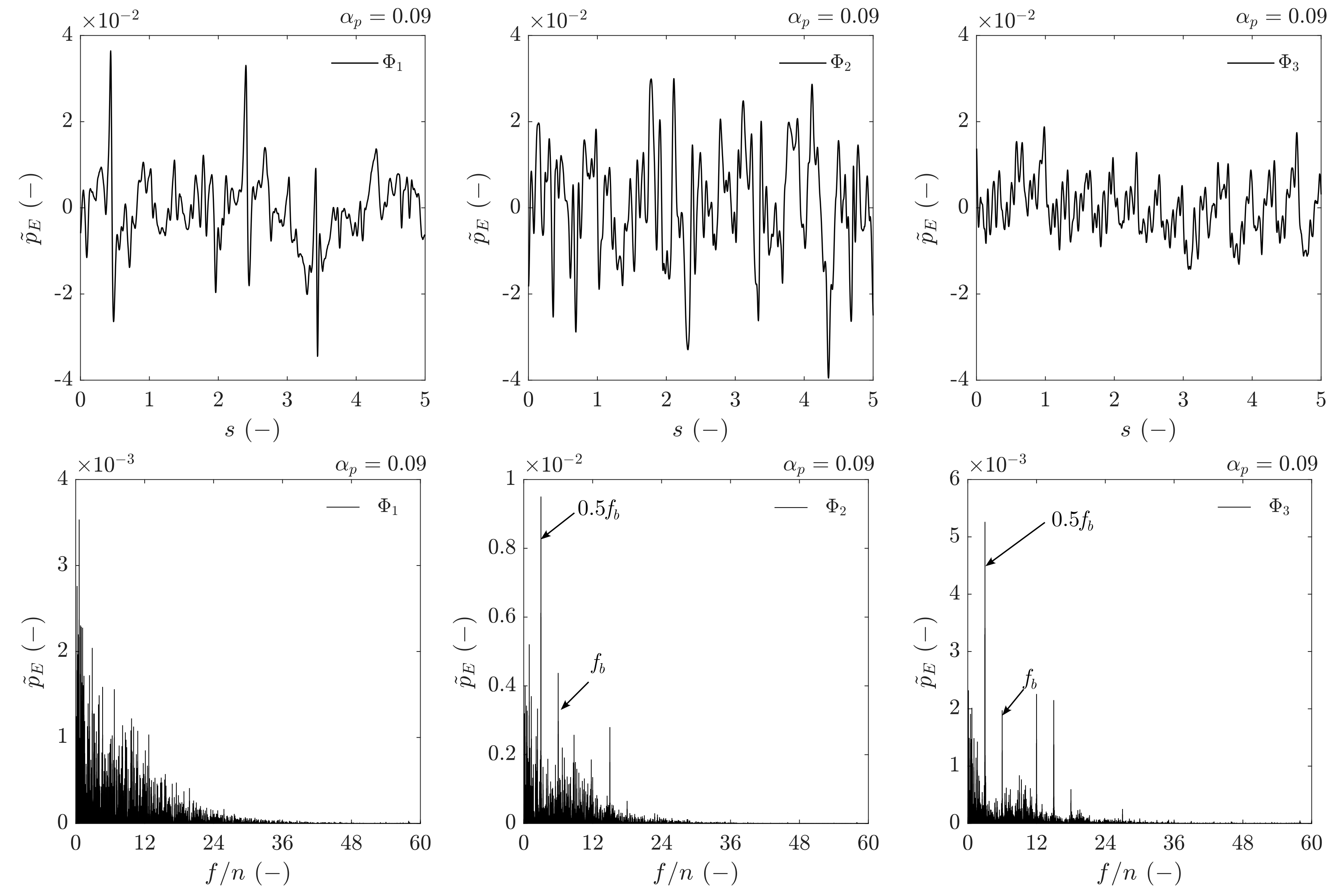}
    \caption{Unsteady pressure fluctuations and frequency spectra in the guide vane passage. The frequency values on x-axis are normalized by the runner angular speed}
    \label{fig:Figure16}
\end{figure}

\begin{figure}[htbp]
    \centering
    \includegraphics[width=1\linewidth]{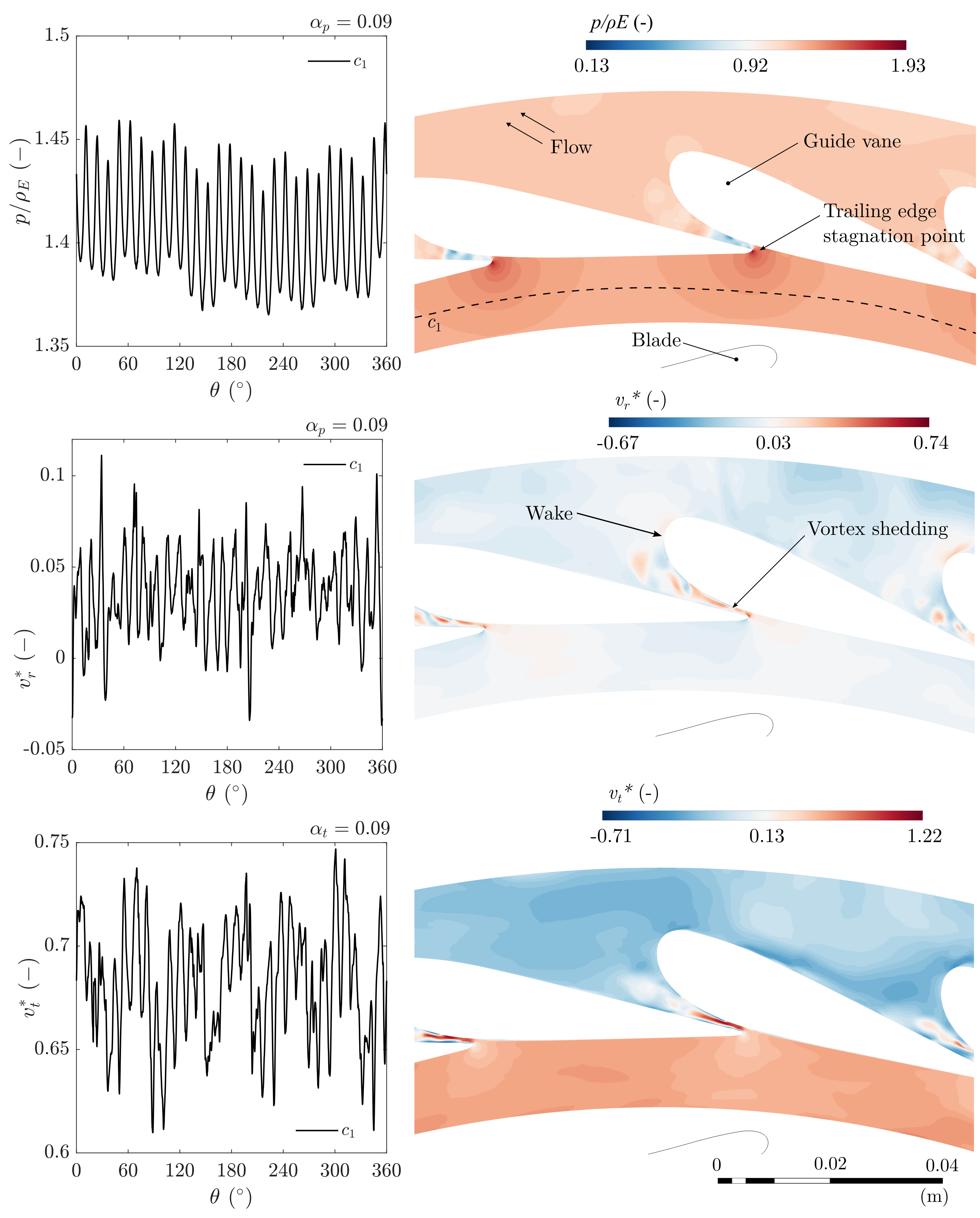}
    \caption{Static pressure and velocity (radial and tangential) variation at the runner inlet vaneless space. The x-axis represents the 0 – 360 degrees of the runner circumference in the vaneless space. Negative radial velocity indicates the vector towards the turbine axis. Negative tangential velocity indicates the vector towards the rotation speed direction of the runner.}
    \label{fig:Figure17}
\end{figure}

When flow leaves runner blade passages, it enters the vaneless space and the guide vane passage. Unsteady vortex shedding from the blade leading edge transports to the vaneless space and interacts with the guide vane’s trailing edge stagnation point. In addition, the effect of the rotor-stator interaction is predominant, and pressure amplitudes vary significantly across the flow path. \autoref{fig:Figure16} provides unsteady pressure fluctuations at three locations, $\Phi1$, $\Phi2$, and $\Phi3$. The fluctuations are shown for the time equals to 5 revolutions of the runner, i.e., $0 - 0.7317$ second. The observation point $\Phi3$ is placed in the vaneless space, as shown in \autoref{fig:Figure3}, and acquires fluctuations, which are resulting from the rotor-stator interactions. Predominant amplitudes correspond to the blade passing frequency $(f_b = 6)$ and its higher harmonics. The frequency with the maximum amplitude corresponds to half of the fundamental frequency, i.e., $0.5f_b$. In addition, wide band frequencies with small amplitudes are obtained, which are related to unsteady vortical flow transporting from the runner. The other observation point $\Phi2$ is placed in the small gap between two guide vanes. Signature of the fluctuations is different from that of the point $\Phi3$. The fluctuations are more stochastic with increased amplitude, though the predominant frequencies correspond to the blade passing and the higher harmonics. The stochastic fluctuations are induced by the unsteady vortex shedding from the trailing edge of the guide vane. The observation point $\Phi1$ is near the guide vane leading edge, in a gap between the stay vane and the guide vane. The fluctuations are largely stochastic, indicating the flow field is highly unstable. When flow leaves the narrow passage of the guide vanes, it rapidly expands—identical to the vaneless space in turbine mode. Moreover, the wake from the guide vane leading edge and the trailing edge vortex shedding combines around this location. This results in complex flow conditions sourced from the adverse pressure gradient on the guide vane.

To investigate the stochastic flow in the guide vane passage, pressure and velocity contours are extracted from the simulation results. \autoref{fig:Figure17} presents the pressure and velocity (radial and tangential) variations in the guide vane passage. The pressure values are normalized by the specific hydraulic energy, and the velocity values are normalized by the characteristic velocity. The pressure and velocity values are acquired along the polyline $(c_1)$ drawn in the vaneless space along the runner circumference; $r* = 1.86$, $z* = 0$, and $\theta = 0 - 360^\circ$. Surprisingly, the pressure variation follows the flow field developed by the guide vane trailing the edge stagnation point. The impact of the runner blade leading edge and the wake appears insignificant. Peak value indicates the guide vane trailing edge stagnation pressure, which can be clearly seen in the pressure contours. Pressure in the narrow gap between two guide vanes is very low, revealing a very high acceleration of flow and the trailing edge vortex transport through the gap. This can be seen in the contours of radial velocity $(v^*_r)$. Flow detaching from the guide vane trailing edge quickly accelerates into the narrow gap, which develops a series of vortex transporting into the gap. Interestingly, the vortex interacts with the wake developed by the leading edge, and the combined vortical flow travels further into the stay vane passage. Contour of tangential velocity $(v^*_t)$ also reveals a very high velocity associated with the trailing edge vortex in the gap. The tangential velocity in the vaneless space is high as it is mainly driven by the angular speed of the runner $(r \times \omega)$.

\section*{Conclusions}
\label{sec:conclusion}
The present study investigated the formation and transport of the unsteady vortex under one of the extreme conditions of a reversible pump-turbine. A computational domain of 120 million nodes is used to conduct large eddy simulations (wall-adapted local eddy-viscosity). In the guide vane passage and the vaneless space, flow characteristics are largely dependent on the rotor-stator interactions. However, the findings showed that there was a longitudinal vortex along the blade span on the leading edge. The longitudinal vortex transports on the high-pressure side of the blade, and splits into two large vortices. Interestingly, they split further downstream and develop the unsteady “string of swirls”. Their spatial and temporal signature appears changing with the blade position. In the draft tube, several instabilities are obtained where the main flow in the center of the draft tube is reversible (pumping) predominantly and flowing towards the runner center. Source of the reversible flow is located at the elbow of the draft tube. In addition, a cascade of swirling zones of secondary flow were obtained in the elbow.

In pump mode, moderate separation of the wall-attached flow was obtained in the draft tube. On the other hand, in the runner, strong flow instabilities were acquired around the trailing edge of the blades. Several vortical zones seemingly developed in the blade passage starting from the trailing edge to the 50\% chord length. Flow around these vortical zones is reversible and transports towards the trailing edge and creates a partial blockade. In the vaneless, surprisingly, the pressure variation follows the flow field developed by the guide vane trailing edge stagnation point. The impact of the runner blade leading edge and the wake appears insignificant. Pressure in the narrow gap between two guide vanes was very low, revealing a very high acceleration of flow and the trailing edge vortex transport through the gap.

Overall, simulation results revealed the formation of “string of swirls” in the runner causing high amplitude stochastic loading and flow instabilities. The high amplitude stochastic loading may potentially develop fatigue on turbine components in both turbine and pump modes.

\section*{Acknowledgements}
The numerical simulations of reversible pump-turbine have been conducted using high performance computing cluster—IDUN at NTNU, Norway.

\section*{Funding}
This research has received financial support from the European Union’s Horizon Europe ``Research and Innovation Action” programme HORIZON-CL5-2023-D3-01-13 under the grant agreement number 101136176. Project: Novel long-term electricity storage technologies for flexible hydropower – store2hydro. 

\section*{Disclaimer}
Funded by the European Union. Views and opinions expressed are however those of the author(s) only and do not necessarily reflect those of the European Union or European Commission. Neither the European Union nor the granting authority can be held responsible for them.

\section*{Conflict of Interest}
The authors have no conflicts to disclose.

\section*{CRediT author contributions}
Chirag Trivedi: Conceptualization; Methodology; Software; Validation; Formal analysis; Investigation; Resources; Data curation; Writing - original draft; Writing - review \& editing; Visualization; Supervision; Project administration; Funding acquisition.

\section*{Data availability statement}
The data that supports the findings of the present study may be provided upon direct request to the author.

\section*{Preprint}
Author’s Original Manuscript (AOM) is submitted to arXiv preprint server \\ (https://arxiv.org/).

\section*{Nomenclature}
\textbf{Variables} \\
$c$ = Chord of the blade (\si{\meter}), circumference \\
$D$ = Runner outlet diameter (\si{\meter}) \\
$E$ = Specific hydraulic energy (\si{\joule\per\kilogram}), $E = gH$ \\
$\hat{e}$ = Error (\si{\percent}) \\
$f$ = Frequency (\si{\hertz}) \\
$H$ = Head (\si{\meter}) \\
$l$ = Length (\si{\meter}) \\
$n$ = Runner angular speed (\si{\per\second}) \\
$n_{ED}$ = Speed factor (-), $nD / \sqrt{E}$ \\
$p$ = Pressure (\si{\pascal}) \\
$p'$ = Fluctuating pressure (\si{\pascal}) \\
$Q$ = Flow rate (\si{\cubic\meter\per\second}) \\
$Q_{ED}$ = Discharge factor (-), $QD^{2}/\sqrt{E}$ \\
$R$ = Radius (\si{\meter}); $R = D/2 = \SI{0.1745}{\meter}$ \\
$s$ = Runner pitch (-) \\
$T$ = Torque (\si{\newton\meter}) \\
$t$ = Time (\si{\second}) \\
$u$ = Velocity (\si{\meter\per\second}) \\
$v$ = Velocity (\si{\meter\per\second}) \\
$v_c$ = Characteristic velocity (\si{\meter\per\second}), $\sqrt{2gH}$ \\
$w$ = Velocity (\si{\meter\per\second}) \\
$Z$ = Number of vanes \\
x, y, z = Spatial coordinates \\[0.75em]

\textbf{Greek Letters} \\
$\alpha$ = Guide vane angle / turbine load (-) \\
$\beta$ = Blade angle (\si{\degree}) \\
$\rho$ = Density (\si{\kilogram\per\cubic\meter}) \\
$\eta$ = Efficiency (-) \\
$\lambda$ = Blade span (-) \\
$\Phi$ = Observation points in guide vane passage \\
$\psi$ = Observation points in draft tube \\
$\Pi$ = Observation points in runner \\[0.75em]

\textbf{Subscripts} \\
b = Blade \\
exp = Experimental \\
ext = Extrapolation \\
gv = Guide vane \\
num = Numerical \\
p = Pump mode \\
r = Radial \\
t = Turbine mode, total, tangential \\
v = Validation \\

\bibliographystyle{asmejour}
\begin{flushleft}
\singlespacing
\bibliography{Bibliography}
\end{flushleft}

\clearpage
\end{document}